\documentclass[10pt, conference, letterpaper]{IEEEtran} 

\AtBeginDocument{%
  \providecommand\BibTeX{{%
    \normalfont B\kern-0.5em{\scshape i\kern-0.25em b}\kern-0.8em\TeX}}}
    
\usepackage[T1]{fontenc}
\usepackage{graphicx}
\usepackage{float}
\usepackage{dsfont}
\usepackage{amsmath}
 
\usepackage[ruled,vlined]{algorithm2e}
\usepackage[font=small]{caption}
\usepackage{subcaption}
\usepackage[export]{adjustbox}
\usepackage{siunitx}
\usepackage{colortbl}
\usepackage{algorithmic}
\usepackage[thinlines]{easytable}
\usepackage{siunitx}
\usepackage[normalem]{ulem}
\usepackage{amsfonts}
\usepackage{multirow,verbatim}
\usepackage[normalem]{ulem}
\usepackage[dvipsnames]{xcolor}
\usepackage{url}
\usepackage{tikz}
\usetikzlibrary{tikzmark}
\usepackage{multirow}
\usepackage{amsthm}
\usepackage[english]{babel}
\usepackage{xcolor}
\usepackage{comment}

\newcommand{\name}{\text{DT-RaDaR}}

\newcommand{\env}{\mathcal{E}}

\newcommand{\reflect}{\rho}

\newcommand{\policy}{\theta}
\newcommand{\qnetwork}{\theta_{Q}}
\newcommand{\targetnetwork}{\theta_{T}}

\newcommand{\state}{s_{t}}
\newcommand{\nextstate}{s_{t+1}}
\newcommand{\action}{a_{t}}

\SetKwComment{Comment}{$\triangleright$\ }{}

\newtheorem{observation}{Observation}
\newtheorem{remark}{Remark}

\SetKwInput{KwInput}{Input}                
\SetKwInput{KwOutput}{Output}              

\begin{document}

\title{$\name$: Digital Twin Assisted Robot Navigation using Differential Ray-Tracing}

\author{
\small
		\IEEEauthorblockN{
        Sunday Amatare\IEEEauthorrefmark{1}, Gaurav Singh\IEEEauthorrefmark{1}, Raul Shakya\IEEEauthorrefmark{1},
        Aavash Kharel\IEEEauthorrefmark{1}, Ahmed Alkhateeb\IEEEauthorrefmark{2} and
        Debashri Roy\IEEEauthorrefmark{1}
        }
\IEEEauthorblockA{The University of Texas at Arlington\IEEEauthorrefmark{1} and
Arizona State University\IEEEauthorrefmark{2}\\
Emails: \{saa3326, gaurav.singh4, rxs6339, axk8168\}@mavs.uta.edu\IEEEauthorrefmark{1}, aalkhateeb@asu.edu\IEEEauthorrefmark{2}, debashri.roy@uta.edu\IEEEauthorrefmark{1}
}
}

\maketitle


\begin{abstract}

Autonomous system navigation is a well-researched and evolving field. Recent advancements in improving robot navigation have sparked increased interest among researchers and practitioners, especially in the use of sensing data. However, this heightened focus has also raised significant privacy concerns, particularly for robots that rely on cameras and LiDAR for navigation. Our innovative concept of Radio Frequency (RF) map generation through ray-tracing (RT) within digital twin environments effectively addresses these concerns. In this paper, we propose $\name$, a robust privacy-preserving, deep reinforcement learning-based framework for robot navigation that leverages RF ray-tracing in both static and dynamic indoor scenarios as well as in smart cities. We introduce a streamlined framework for generating RF digital twins using open-source tools like Blender and NVIDIA's Sionna RT. This approach allows for high-fidelity replication of real-world environments and RF propagation models, optimized for service robot navigation. Several experimental validations and results demonstrate the feasibility of the proposed framework in indoor environments and smart cities, positioning our work as a significant advancement toward the practical implementation of robot navigation using ray-tracing-generated data. 

\end{abstract}


\vspace*{-5pt}
\section{Introduction}\vspace{-0.05in}
\label{sec:intro}


The domain of autonomous robots is expanding swiftly on a global scale, with a broadening range of new uses and increasing interests from established sectors such as automotive, freight, public transit, industrial operations, and defense. One area where autonomous robots have seen increased interest in recent years is e-commerce. This is largely attributed to traffic congestion, especially in urban areas, and the difficulties companies encounter with tight schedules for picking up and delivering items~\cite{ADR}. Additionally, the recent increase in package deliveries, as more people shop online instead of visiting physical stores, has further intensified the demands on human workers. Particularly, autonomous delivery robots (ADR) have become a practical solution for these tasks. These robots can assist human workers in areas where their capabilities are constrained, particularly by helping to reduce their workload. According to Facts and Factors, delivery robot market is projected to reach \$55 billion by 2026, with an annual growth rate of 20.4\%~\cite{ADR}. Additionally, certain companies assert that these robots can lower delivery expenses by as much as 90\%. Companies like Panasonic, Starship, and Robomart are leading the development of ADRs~\cite{ADR}. In addition, major delivery services such as Amazon, FedEx and Uber are utilizing ADRs to improve their delivery operations~\cite{srinivas2022autonomous}.

 \begin{figure}[t!]
     \centering  
       \includegraphics[width=1\linewidth]{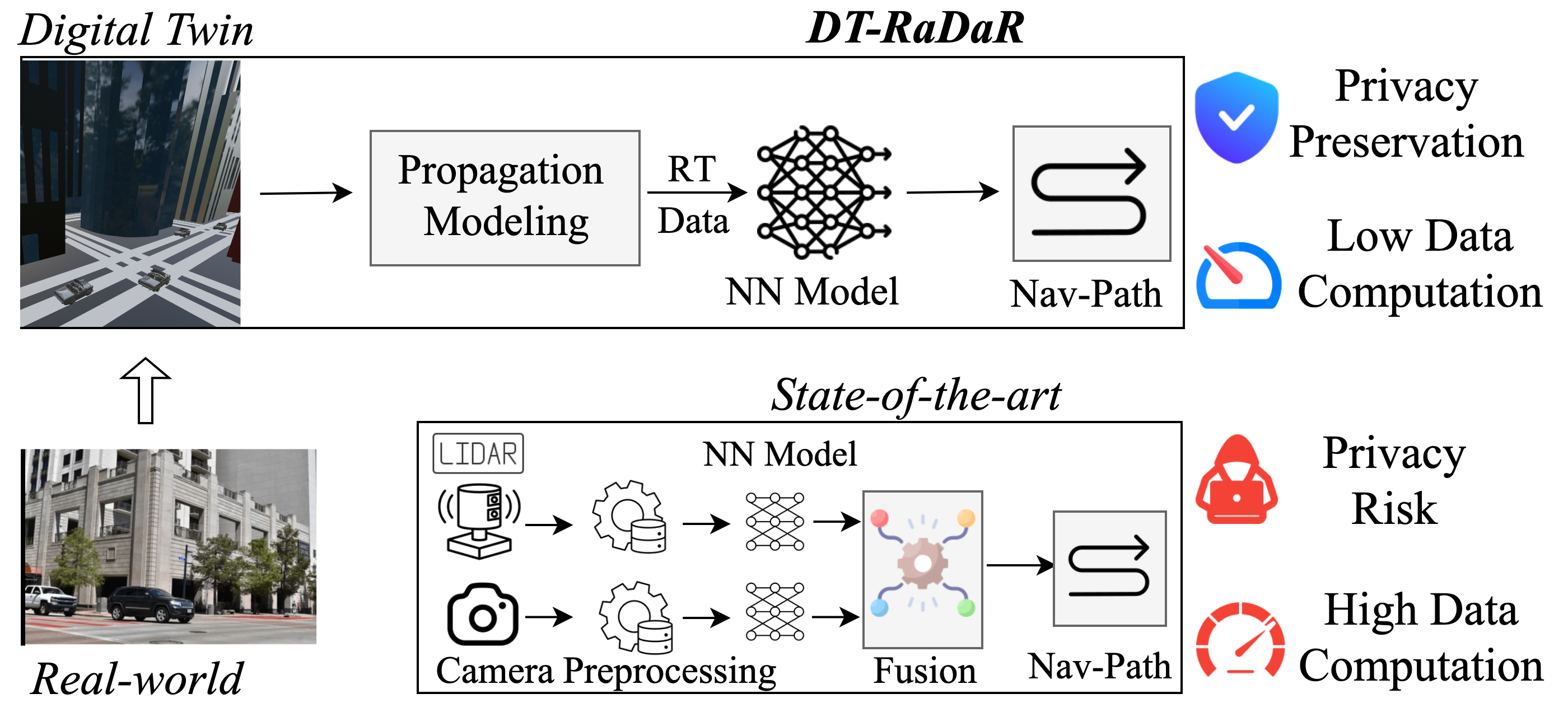}
      \caption{Comparison of sensor-based navigation and $\name$. We show that the $\name$ is privacy preserving and data efficient.}
       \vspace{-0.3in}
      \label{fig:intro-fig}
 \end{figure}

\noindent
{\bf Sensor-enabled Robot Navigation.} Sensing data plays a crucial role in autonomous systems, especially for service robots and ADRs engaged in tasks such as mapping, localization, and navigation within unstructured environments. Among the laser-based solutions for mobile robot navigation and localization, 2D light detection and ranging (LiDAR) sensors stand out as a popular option due to their accessibility and trusted performance. Although, a significant limitation of the 2D LiDAR sensor is its dependence on a single horizontal scan line for sensing, which restricts its capacity to gather detailed spatial data and hinders effective navigation in complex terrains. On the other hand, 3D LiDAR enables a more detailed understanding of the environment by collecting spatial data in multiple dimensions. Yet, its increased expense can be a barrier, discouraging its use in certain applications. In recent years, depth cameras have gained popularity primarily due to their affordability, accessibility, and ability to capture detailed depth information at a lower cost~\cite{camplani2017multiple}. These features make depth cameras a competitive option for various applications, particularly in situations where more expensive solutions, like 3D LiDAR, are impractical. However, depth cameras have limitations in range and accuracy.

\noindent
{\bf Concerns with Sensor-enabled Robot Navigation.} The integration of sensory information into service robots and ADRs presents significant privacy challenges that require careful consideration. Key concerns include the risk of unauthorized monitoring, where individuals may be surveilled without their consent, leading to potential violations of personal privacy. Moreover, the possibility of data breaches raises alarms, as sensitive information collected by these robots could be compromised, exposing users to various security threats. These concerns underscore the necessity of implementing a privacy-preserving system to protect sensitive information collected by service robots. This system should prioritize ethical standards, such as obtaining explicit user consent and upholding transparent data usage policies throughout both the development and deployment stages of these technologies. By doing so, it fosters trust among users and stakeholders, as highlighted in \cite{gatesichapakorn2019ros}. Furthermore, it is essential to recognize that these sensors require significant computational resources due to their need to provide high precision to capture the environmental characteristics critical to effective robot navigation~\cite{gatesichapakorn2019ros}.

\noindent
{\bf Privacy Preservation through Ray-tracing (RT).} The concept of rays is easy to grasp based on our everyday encounters with sunlight. When sunlight passes through a relatively large opening in a wall and enters a room, we can observe the `ray' traveling in a straight path~\cite{yun2015ray}. RF ray-tracing is an established technology that can effectively model the propagation of electromagnetic waves across various relevant scenarios. It has become an effective approach for selectively capturing and analyzing visual and spatial data from the environment while ensuring privacy protection. Ray-tracing models electromagnetic waves as distinct rays that travel in straight lines. When these rays interact with objects in their environment, they can generate various phenomena, such as reflection, refraction, diffraction, and scattering. Differential ray-tracing (DRT) is a specialized extension of conventional ray-tracing that detects slight alterations in the environment, allowing for precise sensitivity analysis ~\cite{hoydis2023sionna,hoydis2023learning}. 
It also identifies essential environmental characteristics required to produce accurate directives for actuation in complex settings, which can be utilized to inform robot navigation decisions. Furthermore, the data generated through DRT includes numerical values of different ray properties, making it a potentially efficient means of representing the environment.

\noindent {\bf Motivation and Contributions of this Paper.} Most state-of-the-art service robots use cameras or LiDAR systems to perceive and interpret their surroundings, enabling efficient navigation and task performance across diverse applications, including autonomous delivery robots. While these technologies enhance the robots' ability to operate in complex environments, they often collect large amounts of visual and spatial data, potentially capturing sensitive information such as individuals' physical characteristics, locations, or behaviors. This poses significant privacy risks, raising concerns about data security, consent, and potential misuse~\cite{song2023calculate, arabo2012privacy}. With this motivation in mind, we propose an innovative RF-based framework, $\name$, aimed at ensuring privacy in robot navigation while advancing technology. The RF-based framework requires robots to be equipped with RF devices, which are often integrated into modern service robots but typically used solely for communication purposes. These RF devices are used for ray-tracing to generate maps, aiding the robots in navigating through unstructured environments. As illustrated in Fig.~\ref{fig:intro-fig}, in $\name$, the novel application of RT for robot navigation requires: (a) Creation of static blueprint digital twins and dynamic digital twins that accurately simulate real-world scenarios, (b) a DRT-based propagation modeling and data generation process within the digital twin, and (c) the implementation of a reinforcement learning (RL) approach that leverages the propagation characteristics of the digital twin to generate optimal navigation paths for the robot. The proposed work requires validation through a specific type of simulation, which includes creating a digital twin from real-world, integrating the digital twin with ray-tracing software, performing propagation characteristics within the imported digital twin, and generating data to train the proposed RL-based robot navigation model. Formally, this paper's contributions are:


\noindent
{\bf {C1. }}  We propose a high-fidelity digital twin creation method that accurately models both indoor and outdoor environments by extracting real-world features through the open-source Blender tool and its OpenStreetMap (OSM) add-on. 

\noindent
{\bf {C2. }}  We propose a methodology for precisely configuring scenes and generating the propagation characteristics of various environments. This approach employs DRT on the digital twin of each scene using NVIDIA's Sionna RT-based software to generate propagation data across all grids within the scene.

\noindent
{\bf {C3. }}  We propose a RL-based algorithm by designing: (a) a customized environment using the propagation data generated from the ray-tracing tool and (b) a customized state extraction process from the data generated using the ray-tracing, four actuation actions for the robot, and a reward function that provides feedback to the (DQN) agent received from the environment. Following the realistic scenario, we train the DQN agent on a blueprint digital twin and perform inference on dynamic digital twins. 

\noindent
{\bf {C4. }} We present a first-of-its-kind (to the best of our knowledge) ray-tracing-based digital twin dataset that replicates both indoor laboratory environments and outdoor smart city scenarios, specifically modeled on downtown Dallas and downtown Houston. These highly realistic digital twins are curated using publicly available software tools, ensuring accuracy and relevance to real-world settings. 

\noindent
{\bf {C5. }}  We validate our overall framework, $\name$, through rigorous experimental validations on the collected datasets. We also conduct experiments where the $\name$ agent is trained on a twin and its inference performance is analyzed on another. We observe that RT-based training requires $\sim < 74.4\%$ of training time of similar methods which uses LiDAR/Camera sensors. Furthermore, upon acceptance of this article, we pledge to release our codebase, digital twin models, ray tracing datasets in \cite{twist}.

\vspace*{-3pt}

\section{Related Works and Motivation}
\label{sec:related_works}


\noindent{\bf Indoor Robot Navigation:}
In this type of navigation, robots employ various techniques and systems to understand their environment, plan paths, and avoid obstacles while moving through spaces such as homes, hospitals, offices, and warehouses. Almeida {\em et al.} \cite{Almeida} propose a precise magnetic mapping method that is reliable and accurate, effectively addressing many indoor localization challenges using magnetic data. Noh {\em et al.} \cite{Noh} present an autonomous mobile robot system that uses open-source ROS packages to address the challenge of avoiding collisions with both static and dynamic objects. Bavle {\em et al.} \cite{Bavle} introduce using odometry readings and planar surfaces extracted from 3D LiDAR scans to construct a Situational Graph for robot pose estimation. Amatare {\em et al.} \cite{Sunday24} introduce an RF-based robot navigation system leveraging Q-Learning in a static environment. Kulhanek {\em et al.} \cite{Jonas} propose a deep reinforcement learning method for visual navigation in real-world settings, based on the parallel advantage actor-critic algorithm, enhanced with auxiliary tasks and curriculum learning.

\noindent{\bf Outdoor Robot Navigation:}
For outdoor navigation the robots utilize advanced technologies and methods to navigate and operate effectively due to the challenges such as varying terrain, dynamic obstacles, and changing weather conditions. Palazzo {\em et al.} \cite{Palazzo} introduce a deep learning-based approach to assess and predict the traversability of various routes within the field of view of an onboard RGB camera. Karnan {\em et al.} \cite{Karnan} propose a method combining imitation learning and inverse reinforcement learning for social navigation in mobile robots. They utilize a socially compliant navigation dataset to address the scarcity of large-scale datasets that capture socially appropriate robot behavior in real-world environments. Elnoor {\em et al.} \cite{Elnoor} present a technique that leverages proprioceptive data to assess terrain traversability in real time for legged robots. Their approach also incorporates sensors to accurately evaluate terrain stability, the risk of entrapment, and potential crashes. Sorokin {\em et al.} \cite{Sorokin} introduce a quadrupedal robot designed to navigate sidewalks by following a route generated by a public map service. The robot uses onboard sensors to steer clear of obstacles and pedestrians, ensuring safe and collision-free movement.

\noindent
{\bf Motivation for Designing $\name$:} 
All the highlighted approaches to robot navigation face challenges related to data privacy and the computational demands associated with the use of high-definition sensors, such as RGB-D cameras and LiDAR systems. Additionally, previous RF-based approach focused solely on indoor, static environment, whereas in practice, scene environments frequently change, both indoors and outdoors, across nearly all applications~\cite{Sunday2024}. In this paper, we propose $\name$, a robust privacy-preserving, deep reinforcement learning-based framework for robot navigation that leverages RF ray-tracing in both static and dynamic indoor scenarios as well as in smart cities.



\vspace*{-3pt}
\section{System Model and Problem Formulation}
\label{sec:problem-formulation}


\subsection{Problem Formulation}
\label{sec:problem}
We consider robot $R$ in the environment with $O$ obstacles. In our setting, the robot is equipped with RF receiver and edge devices. We have a transmitter in the environment which is transmitting in the 2.4GHz band. The RF receivers listen to the 2.4GHz band and perform ray-tracing. Each edge device of the robot runs a decision algorithm from the ray-tracing data and generates the next coordinate for navigating through the environment~(see Fig.~\ref{fig:intro-fig}). The path planning policy targets to minimize the number of steps reaching to the target, while avoiding the obstacles. We use deep reinforcement learning algorithm to find the optimum policy that meets the above requirements. Thus, given a path planning policy $\theta$, we formulate our objective as: 
\begin{subequations}
\label{eq:optimization_problem_formulation}
\vspace{-0.1in}
  \begin{align}
    &\underset{\policy}{\text{Maximize:}}\quad \underset{\theta}{\mathbb{E}} \bigg[\sum_{t=0}^{T-1}f_{\mathcal{R}}^{\theta}(\state, \action)\bigg],\\
    &\text{s.t\quad} \action = \policy(\state),\\
&~~~~~\sum_{t=1}^{\mathcal{T}}\varphi_t = 0.
    \end{align}
    \vspace{-0.1in}
\end{subequations}

Here, $\state$ and $\action$, denote the state and action for the robot~$R$ at step $t$ ($T$ total steps). The action space includes movement either in the X or in the Y directions. Moreover, $f_{\mathcal{R}}^{\theta}$ denotes a reward function, which is a function of state and action. Finally, $\varphi_t$ is a Boolean predicate, with $\varphi_t$ to be 1 if collision happens, and 0 otherwise, that ensures avoiding collisions.\vspace{-0.05in}

\begin{figure*}
    \centering
    \includegraphics[width=\linewidth, clip]{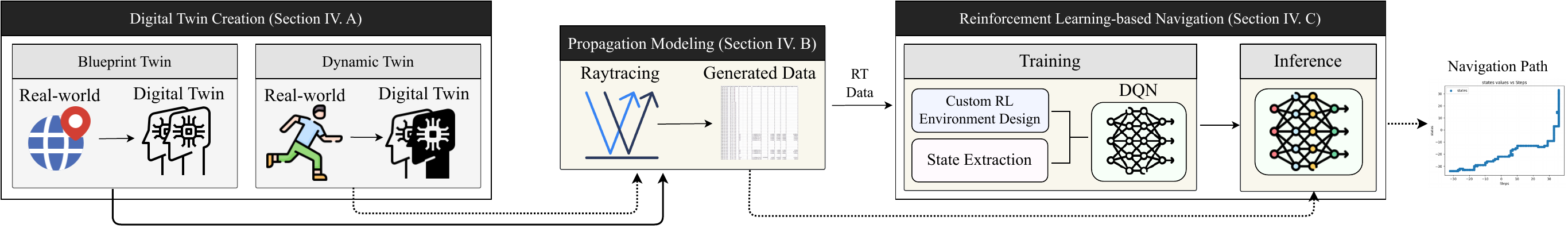}
    \caption{Overview of the proposed system architecture. $\name$ continuously monitors the environment. In {\em Module 1}, we generate the digital twins. The re-training is performed over an interval~({\em Module 2}), we update the {\em Module 3} which eventually updates path planning policy by  optimizing the navigation~({\em Module 3}). The solid line shows the training path, and dashed line represent inference path.}
        \vspace{-15pt}
    \label{fig:system_archi}
\end{figure*}

\subsection{System Architecture in \name}\vspace{-0.05in}
\label{sec:system_model}
An overall view of our framework is shown in Fig.~\ref{fig:system_archi} and consists of three main modules as follows:

\begin{itemize}
    \item \textbf{Digital Twin Creation} {\em (Module 1)}{\bf :} We create a digital twin by directly extracting the real world features to digital world via sensors. The training of RL algorithm is done over a {\em blueprint digital twin}, and the inference and deployment is conducted over the real-time twins, named as {\em dynamic digital twin}. While the transmitter is placed at the roof area, the receiver (robot) is placed on the floor of the room alongside obstacles~(details in Sec.~\ref{Sec:proposed_mone}).
    
    \item \textbf{Propagation Modeling} {\em (Module 2)}{\bf :} The autonomous robot receives the instantaneous local ray-tracing data from the transmitting device in the digital twin and acts appropriately to detect the obstacles in close and distant ranges~(details in Sec.~\ref{Sec:proposed_mtwo}).
    
    \item \textbf{RL-based Navigation} {\em (Module 3)}{\bf :} We design a  RL-based algorithm that leverages the ray-tracing data and the propagation characteristics from the digital twin for the robot navigation~(details in Sec.~\ref{Sec:proposed_mthree}).

\end{itemize}

\begin{remark}
Our digital twin generation method is designed to generalize to realistic scenarios by utilizing training data collected from an initial blueprint digital twin. As the blueprint digital twin is continuously updated over time, re-training is conducted to ensure model accuracy and adaptability.
\end{remark}
\vspace*{-3pt}

\section{$\name$ Framework}
\label{sec:proposed_framework}

\subsection{Module 1: Digital Twin Creation}\vspace{-0.03in}
\label{Sec:proposed_mone}

In the $\name$, we consider properties with respect to map precision and radio propagation properties. While we integrate several specific key metrics in the $\name$, our baseline can be extended in the future to different environmental setup. We initialize the created digital twin as $\env$ = $f(\mathrm{map},\mathrm{O}, \reflect)$. Here, $\mathrm{map}$ and $\mathrm{O}$ denote the imported Blender~\cite{blender} map and present structures or obstacles for the twin $\env$, and $\reflect$ are number of allowed reflections for the  created twin.  We characterize objects present in the twin with 
$\mathrm{O}_{u} = \{(x_k, y_k, d_{0,k}, d_{1,k}, d_{2,k})\}_{k=1}^{N^{\text{obs}}},$
where $N^{\text{obs}}$ is the number of structures or obstacles that are present in the twin $\env$. Moreover, ($x_k$, $y_k$) are the obstacle coordinates on the map, whereas $d_{0,k}$, $d_{1,k}$, $d_{2,k}$ represent the height, width, and length of the obstacle-$k$. 

As mentioned in Sec.~\ref{sec:system_model}, the digital twins are categorized as (a) blueprint digital twin which provides the blueprint of the scenario with static objects for the RL agent to train on and (b) dynamic digital twin which captures the real-time changes within the scenario during inference.


\subsection{Module 2: Propagation Modeling}
\vspace{-0.03in}
\label{Sec:proposed_mtwo}
We use the open source Sionna RT~\cite{nvidia_rt_details} tool to generate the propagation characteristics of the created digital twin $\env$ by employing RF ray-tracing. For a given transmitter $TX$, a propagation map is a rectangular surface with arbitrary orientation subdivided into rectangular cells of size $|\mathcal{C}|$. A parameter $\eta$ controls the granularity of the map. The propagation map associates with every cell $(\mathcal{C}_i, \mathcal{C}_j)$. In $\name$, for every ray $n$ intersecting the propagation map cell $(\mathcal{C}_i, \mathcal{C}_j)$, the channel coefficients $\Theta$, phase shifts $\phi$, the azimuth angles of departure $\alpha_d$ and arrival $\alpha_a$, zenith angles of departure $\zeta_d$ and arrival $\zeta_a$ are computed.\vspace{-0.05in}  

\subsection{Module 3: RL-based Navigation (Training)}
\begin{figure}
    \centering
    \vspace{-0.15in}
    \includegraphics[trim={0cm 3.2cm 0 0},clip,width=0.75\linewidth]{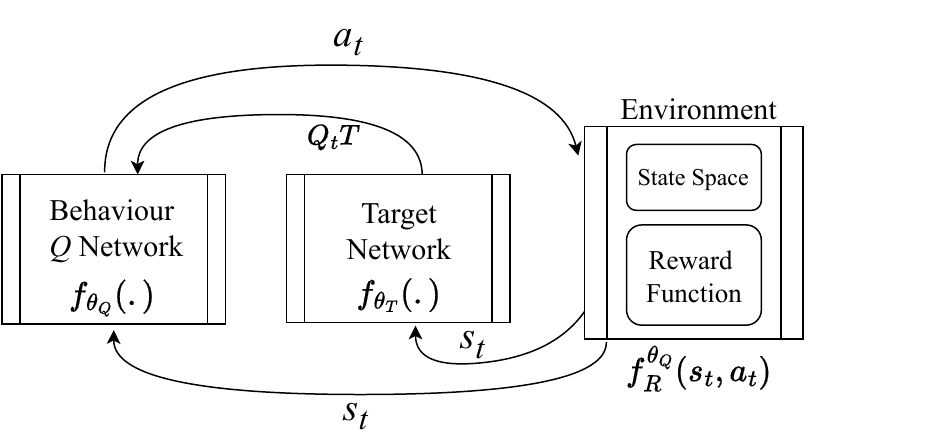}
    \caption{Overview of the proposed DQN agent. We formulate the state and reward space to optimize navigation.}\vspace{-0.2in}
    \label{fig:q-learning-agent}
\end{figure}

\vspace{-0.03in}
\label{Sec:proposed_mthree}
Once the propagation modeling is performed for the digital twin $\env$, we use  reinforcement learning algorithm to solve Eq.~\ref{eq:optimization_problem_formulation} and obtain the optimum policy for navigating robot $R$. 
We use the Deep Q-Network (DQN) based training for performing the reinforcement learning to identify the policy for optimum path planning as shown in Fig.~\ref{fig:q-learning-agent}. The agent takes as input the state array $\state$ and outputs the action $\action$ following the policy~$\policy$.

\subsubsection{Custom RL Environment Design} 
An RL environment represents the context in which an RL agent functions, serving as a simulated model with which the agent interacts by executing actions, receiving rewards or penalties, and transitioning to new states accordingly. This environment furnishes the agent with its initial state and updates it to the next state following each action taken by the agent, while also providing the reward back to the agent received at each new state. However, given the novelty of our proposed RT-based robot navigation approach, custom RL environment must be generated to accurately capture the propagation characteristics of the digital twin $\env$. 
As depicted in Fig.~\ref{fig:q-learning-agent}, the custom RL environment encapsulates both state space and reward function which are further elaborated in the subsequent sections.


\subsubsection{State Extraction}
\label{subsubsec:state}
The state space is extracted from the {Module 2}. We define the state space $\state$ of the robot $R$ at step $t$ as: $\state = \{R_x^{t}, R_y^{t}, \zeta^t_d, \alpha^t_d, \zeta^t_a, \alpha^t_a, |\Theta|^t, \phi^t\}$, where $R_x^{t}$ is the X coordinate of the robot location on the map, $R_y^{t}$ is the Y coordinate of the robot location on the map, $\zeta_d^t$ is zenith angle of departure, $\alpha_d^t$ is azimuth angle of departure, $\zeta_a^t$ is zenith angle of arrival, $\alpha_a^t$ is azimuth angle of arrival, $|\Theta|^t$ is channel coefficient magnitude, and $\phi^t$ is phase shift at step $t$. The overall state array has $8$ elements, encapsulating both RF propagation characteristics and navigation elements.

\subsubsection{DQN Agent Design}
Given the state array at the timestep $t$, the DQN agent is trained by interleaving optimization of two neural networks: (a) one neural network which estimates the actions of the agent using the current state based on the Q values and trains them is named as {\em behaviour Q network}, denoted as $f_{\qnetwork}(.)$ with $\qnetwork$ as {\em behavior policy}; and (b) another neural network which sets the target outputs for the behavior Q network to be trained on is named as {\em target Q network}, denoted as $f_{\targetnetwork}(.)$ with $\targetnetwork$ as {\em target policy}.

\noindent{\bf Behavior Q Network.} The behavior Q network policy~($\qnetwork$) is generated by training the behavior Q network which estimates the Q values as: $Q^{Q}_t = f_{\qnetwork}(\state)$. 

\noindent{\bf Action Space.} The action is taken by following the $\epsilon$-greedy policy, $\action=\epsilon-\text{greedy}(Q^{Q}_t)$. The action $\action = \{X-, X+, Y-, Y+\}$, representing traversing either in X direction ($X-$ and $X+$) or Y direction ($Y-$ and $Y+$) to reach the next state's X and Y coordinate ($R_x^{t+1}, R_y^{t+1}$) by the robot $R$ at timestep $(t+1)$.

\noindent{\bf Reward Function.} The reward is calculated based on the next state reached by the robot by taking the action generated by the behavior deep Q network (DQN) in the current state. We define the overall reward as (details are in Table~\ref{tab:reward_description}):
$
    f_{R}^{\qnetwork}(\state, \action) = r_{d,d^{'}}^{t} + r_{arr}^{t} + r_{obs}^{t}.$
Here, $\state$ and $\action$ denote the state array and action of the robot $R$ respectively at timestep $t$. The reward is computed following the policy $\qnetwork$, encapsulating the DQN learning agent. Intuitively, the relative distance reward~($r_{d,d^{'}}^{t}$) is increased when the robot gets closer to the target (target coordinates: $(Tg^t_x, Tg^t_y)$ at timestep $t$) and vice versa. On reaching the goal state the robot achieves the reward of $5K$. On the other hand ($r_{obs}^{t}$) punishes the robot when colliding with any obstacle with a reward of $-5K$.

\noindent{\bf Target Network.} The target network policy~($\targetnetwork$) is defined using a feed forward network called as target network which takes the next state $\nextstate$ as input and estimates the target Q values for training the behavior Q network. The target Q values are denoted as: $Q^{T}_t = f_{\targetnetwork}(\nextstate)$. The rewards received by traversing to the next state is also added to the target Q values discounted by a factor of gamma representing the future sum of rewards, eventually: $Q^{T}_t = Q^{T}_t + f_{R}^{\qnetwork}(\state, \action)$.




\begin{table}[!t]
 \caption{The reward function of the proposed DQN agent.}
\resizebox{1\linewidth}{!}{
\begin{tabular}{|c|c|c|}
\hline
\textbf{Symbol}  & \textbf{Reward Description} &   \textbf{Reward Value}\\
\hline
\hline
$r_{d,d^{'}}^{t}$ & Relative distance to target & $-(|Tg^t_x-R^t_x|^2 + |Tg^t_y-R^t_y|^2)$\\
\hline
$r_{arr}^{t}$ & Reaching target & 5000\\
\hline
\multirow{3}{*}{$r_{obs}^{t}$} & \multirow{3}{*}{Distance to obstacle } & -5000 \hspace{0.4mm} (if $(|Tg^t_x-R^t_x|^2 + |Tg^t_y-R^t_y|) = 0$) \\
& & 0 \hspace{0.4mm} (otherwise) \\
\hline
\end{tabular}}
\vspace{-0.2in}
 \label{tab:reward_description}
\end{table}


\noindent{\bf Model Training on Blueprint Digital Twins.} At the training phase, the DQN agent interacts with the environment by passing the current state $\state$ to the behavior Q network $f_{\qnetwork}(.)$ which generates the action traversing the agent to the next state $\nextstate$ using the behavior Q values $Q^{Q}_t$ following the $\epsilon$-greedy policy. This next state $\nextstate$ is passed to the target network $f_{\targetnetwork}(.)$ which generates the target Q values $Q^{T}_{t+1}$, which works as target values after discounting them and adding reward to them, for the behavior Q network. Hence we design a custom loss function $\mathcal{L}(.)$ using the Bellman equations~\cite{10.5555/3312046} and  a discount factor of $\gamma$.  The loss function $\mathcal{L}(.)$  considers $Q^{Q}_t$ as current Q value and  $Q^{T}_{t+1}$ as target Q value.  The $\mathcal{L}(.)$ uses Huber loss which acts like the mean squared error when the error is small, but like the mean absolute error when the error is large. Formally, 

\begin{equation}  
\label{eqn:huber_loss}
\vspace{-0.1in}
    \begin{split}
    \mathcal{L}(Q^{Q}_t, Q^{T}_{t+1}) &= 
    \begin{cases}
       \dfrac{1}{2}(Q^{T}_{t+1} - Q^{Q}_t)^2 &  for |Q^{T}_{t+1} - Q^{Q}_t|\leq 1 \\
       |Q^{T}_{t+1} - Q^{Q}_t)^2| - \dfrac{1}{2} & \text{Otherwise}\\
     \end{cases}
    \end{split}
\end{equation}

Depending on the loss, the behavior model weights are updated through back propagation. After certain number of training iterations, the model weights of the behavior Q network is copied to the target Q network so that the target Q values are generated using the updated trained network.

\subsection{Module 3: RL-based Navigation (Inference)}\vspace{-0.05in}
\label{Sec:proposed_mfour}
During the inference, the states are extracted by following {\em Module 2} (Sec.~\ref{Sec:proposed_mtwo}), denoted as: $s^{d}_t$ (details in Sec.~\ref{subsubsec:state}). 

\noindent{\bf Model Inference on both Blueprint and Dynamic Digital Twins.} The generated action during inference: $a^{d}_t = \theta (s^{d}_t)$, where $a^{d}_t \in \{X-, X+, Y-, Y+ \}$ and $\theta$ is the trained model from {\em Module 3}, Sec.~\ref{Sec:proposed_mthree}. The inference can be applied to both the blueprint and dynamic digital twins.



\vspace*{-3pt}
\section{$\name$ Dataset}
\label{sec:testbed}
\vspace{-0.03in}
In this section we introduce, to the best of our knowledge, a pioneering ray-tracing-based digital twin dataset that accurately replicates both indoor laboratory environments and outdoor smart city scenarios, specifically modeled after downtown Dallas and downtown Houston. These meticulously crafted digital twins, developed using publicly available software tools, offer high realism and relevance to real-world conditions. Next, we provide step-by-step details of the dataset generation process for enabling the research community to replicate and further explore its applications.

\begin{figure}
    \centering
    \includegraphics[width=\linewidth]{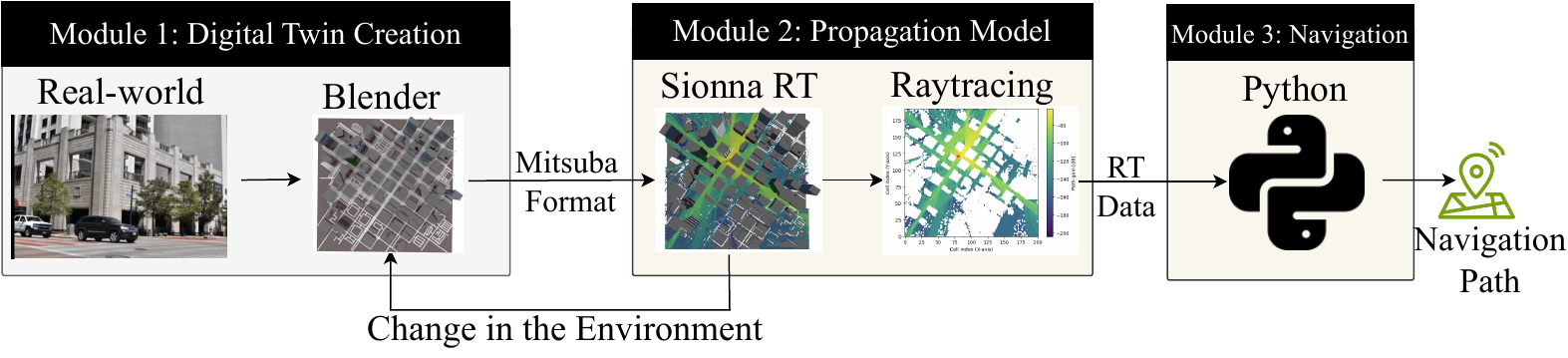} 
    \vspace{-18pt}
    \caption{Overview of various used tools for implementing $\name$.}
\label{fig:system_overview}
    \vspace{-15pt}
\end{figure}

\noindent
{\bf Experimental Platform.} We collect the $\name$ dataset using: (a) an Intel i7 system with Ubuntu, Blender LTS v3.6.12~\cite{blender}, and NVIDIA's Sionna RT~\cite{hoydis2023sionna} for {\em Module 1} and {\em 2}. The dataset generation includes digital twin creation, subsequent ray-tracing and propagation characterization using Sionna RT, as shown in the first two blocks of Fig.~\ref{fig:system_overview}. Details about the robot navigation on the collected data is presented in subsequent section Sec.~\ref{sec:performance}.
\vspace{-0.1in}

\subsection{Experimental Scenarios}
\label{Sec:scenario}
\vspace{-0.03in}
We consider four distinct scenarios, encompassing two indoor and two outdoor environments. The indoor environments include a laboratory cubicle area and our lab's meeting space, while the outdoor environments are represented by smart cities: downtown Dallas and downtown Houston. The indoor scenes include office equipments like tables, chairs, TVs, computers, walls, and floors. The material properties of these objects are represented as wood, metal, glass, marble, or concrete. We present the features of the physical world for each of the experimental scenarios.

\noindent $\bullet$ {\bf Indoor. Lab Cubicle:} The lab cubicle area features 7 metal chairs that serve as student workstations, along with a table surrounded by 4 additional chairs for dining and relaxation. The representative physical world is shown in Fig.~\ref{fig:blueprint_twin} (a).

\noindent $\bullet$ {\bf Indoor. Lab Meeting:} The lab meeting area includes an open space with various objects towards the wall of the meeting room. We have a wooden table and circulating 8 metal chairs in the middle of the room, shown in Fig.~\ref{fig:blueprint_twin} (d). 

\noindent $\bullet$ {\bf Outdoor. Downtown Dallas:} Downtown Dallas measuring 0.4 by 0.5 km, with coordinates [32.78243, -96.80136] for the northwest and [32.77759, -96.79671] for the southeast. The scene includes 37 buildings, along with roads and 10 parking lots, with object materials represented as wood, metal, brick, concrete, or marble. The considered physical area is shown in Fig.~\ref{fig:blueprint_twin} (g) in map form. 

\noindent $\bullet$ {\bf Outdoor. Downtown Houston:} For Downtown Houston, we choose an area measuring 0.8 by 0.8 km, with coordinates [29.75959, -95.37055] for the northwest and [29.75249, -95.36208] for the southeast. This scene includes 58 buildings, 29 parking lots, and numerous roads, with materials represented as wood, metal, glass, marble, brick, or concrete. The considered map is shown in Fig.~\ref{fig:blueprint_twin} (j).\vspace{-0.05in}

\subsection{Digital Twin Creation}\vspace{-0.05in}
\label{subsec:testbed_digital_twin_creation}
We create the digital twin from our physical indoor laboratory environment, as shown in the first block of Fig.~\ref{fig:system_overview}. By harnessing Blender and its OpenStreetMap (OSM) add-on, we digitally recreate an accurate indoor environment from the real world, including scene objects with various radio material properties. The inclusion of obstacles in the digital twin is important for providing information about the environment and guiding path planning and decision-making. This is because effective obstacle detection, avoidance, and understanding are essential capabilities for autonomous robots operating in diverse and dynamic environments. This digital twin is exported from Blender in {\tt Mitsuba 3} {\tt .xml} file format which is responsible for rendering the scene and importation to Sionna RT~\cite{hoydis2023sionna} for propagation modeling. In Sionna, a set of international telecommunication union (ITU) materials with corresponding radio properties~\cite{hoydis2023sionna} is available for each object, ensuring realism and compatibility. Note that the utilization of Blender is solely for the digital twin generation and is not involved in making decisions for robot navigation. More details about the digital twin creation are presented in our previous work~\cite{Sunday2024,Sunday24,Amat2024}.

\subsubsection{Blueprint Digital Twins}
\label{subsubsec:testbed_blueprint_twin}
We generate a blueprint digital twin that incorporates only {\em static} objects. The intuition behind generating this set of twins is to provide the RL agent the blueprint data of an environment to get trained on.  Various properties of the such designed digital twins are:

\noindent $\bullet$ {\bf Lab Cubicle Blueprint Digital Twin.} In this digital twin, we model 11 chairs, 7 cubicles, 1 table and wall replicating the physical world. The carpet floor is represented with \textit{ITU-concrete}, the wall with \textit{ITU-brick}, the chairs with \textit{ITU-wood}, and the cubicles with \textit{ITU-metal} in Blender. The generated twin is shown in Fig.~\ref{fig:blueprint_twin} (b).

\noindent $\bullet$ {\bf Lab Meeting Blueprint Digital Twin.} For our lab meeting space, we model 10 chairs, 4 tables, 1 TV, 2 monitors, 1 board, 2 racks, walls, and carpet to simulate the real world. In Blender, chairs, tables, boards, and racks are modeled with \textit{ITU-wood}, the TV and monitors with \textit{ITU-metal}, the walls with \textit{ITU-brick}, and the carpet with \textit{ITU-concrete}. The twin representation is in Fig.~\ref{fig:blueprint_twin} (e).

\noindent $\bullet$ {\bf Downtown Dallas Blueprint Digital Twin.} For the digital twin creation of downtown Dallas, we model the measured 0.4 by 0.5 km area consisting of 37 buildings, 10 parking lots, and numerous roads to replicate real-world characteristics. In Blender, the buildings are modeled with their physical properties using \textit{ITU-brick}, \textit{ITU-marble}, \textit{ITU-concrete}, or \textit{ITU-glass}. The parking lots and roads are modeled with \textit{ITU-concrete}. The twin is shown in Fig.~\ref{fig:blueprint_twin} (h).


\noindent $\bullet$ {\bf Downtown Houston Blueprint Digital Twin.} We model the measured 0.8 by 0.8 km area of downtown Dallas, encompassing 58 buildings, 29 parking lots, and various roads to replicate the physical properties, shown in Fig.~\ref{fig:blueprint_twin} (k). In Blender, the buildings are modeled with their physical properties using \textit{ITU-brick}, \textit{ITU-marble}, \textit{ITU-concrete}, or \textit{ITU-glass}. The parking lots and roads are modeled with \textit{ITU-concrete}. 




\subsubsection{Dynamic Digital Twins}
\label{subsubsec:testbed_dynamic_twin}
\begin{figure}
	\centering
	\begin{subfigure}{0.4\linewidth}
        \centering
        \includegraphics[width=\linewidth]{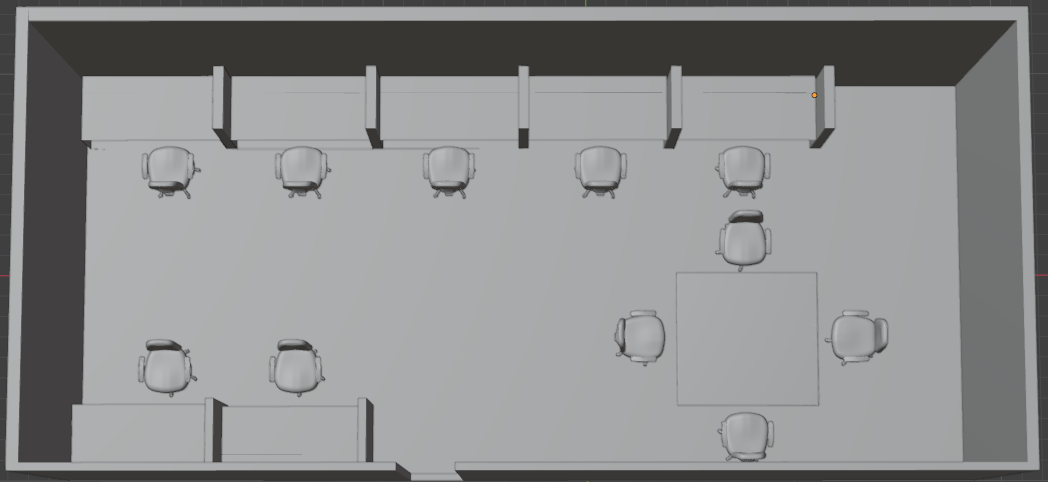}
		\caption{Blueprint Lab Cubicle Scene}
	\end{subfigure}
	\begin{subfigure}{0.4\linewidth}
        \centering
        \includegraphics[width=\linewidth]{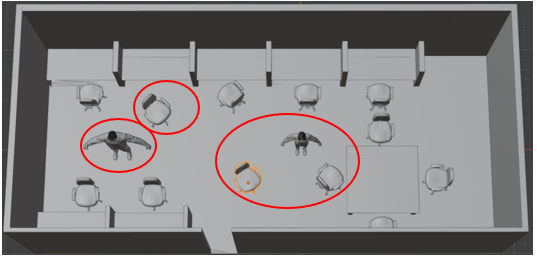}
        		\caption{Dynamic Lab Cubicle Scene}
	\end{subfigure} 
 	\begin{subfigure}{0.4\linewidth}
        \centering
        \includegraphics[width=\linewidth]{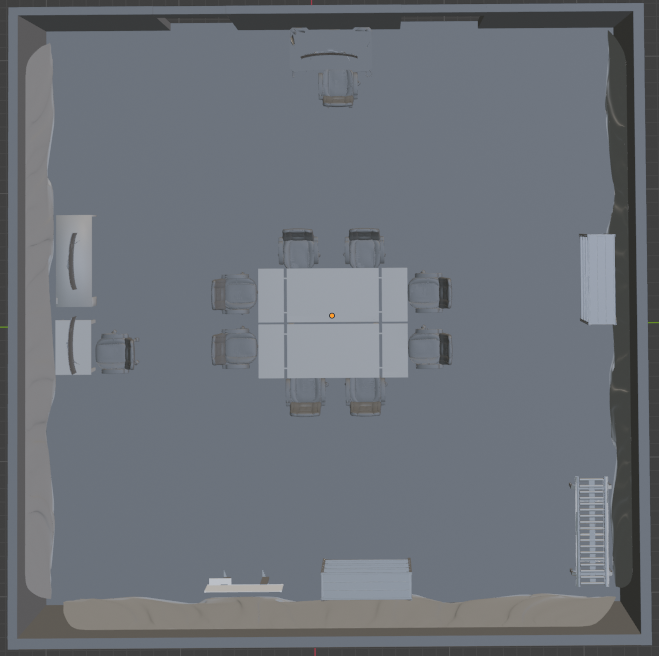}
		\caption{Blueprint Lab Meet Scene}
	\end{subfigure}
	\begin{subfigure}{0.4\linewidth}
        \centering
        \includegraphics[width=\linewidth]{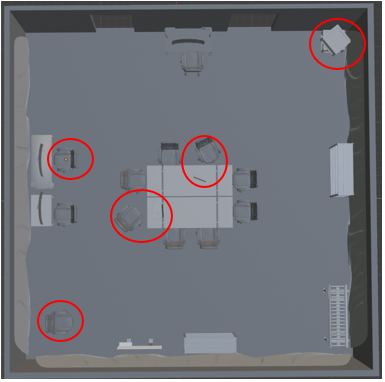}
        		\caption{Dynamic Lab Meet Scene}
	\end{subfigure} 
 	\begin{subfigure}{0.4\linewidth}
        \centering
        \includegraphics[width=\linewidth]{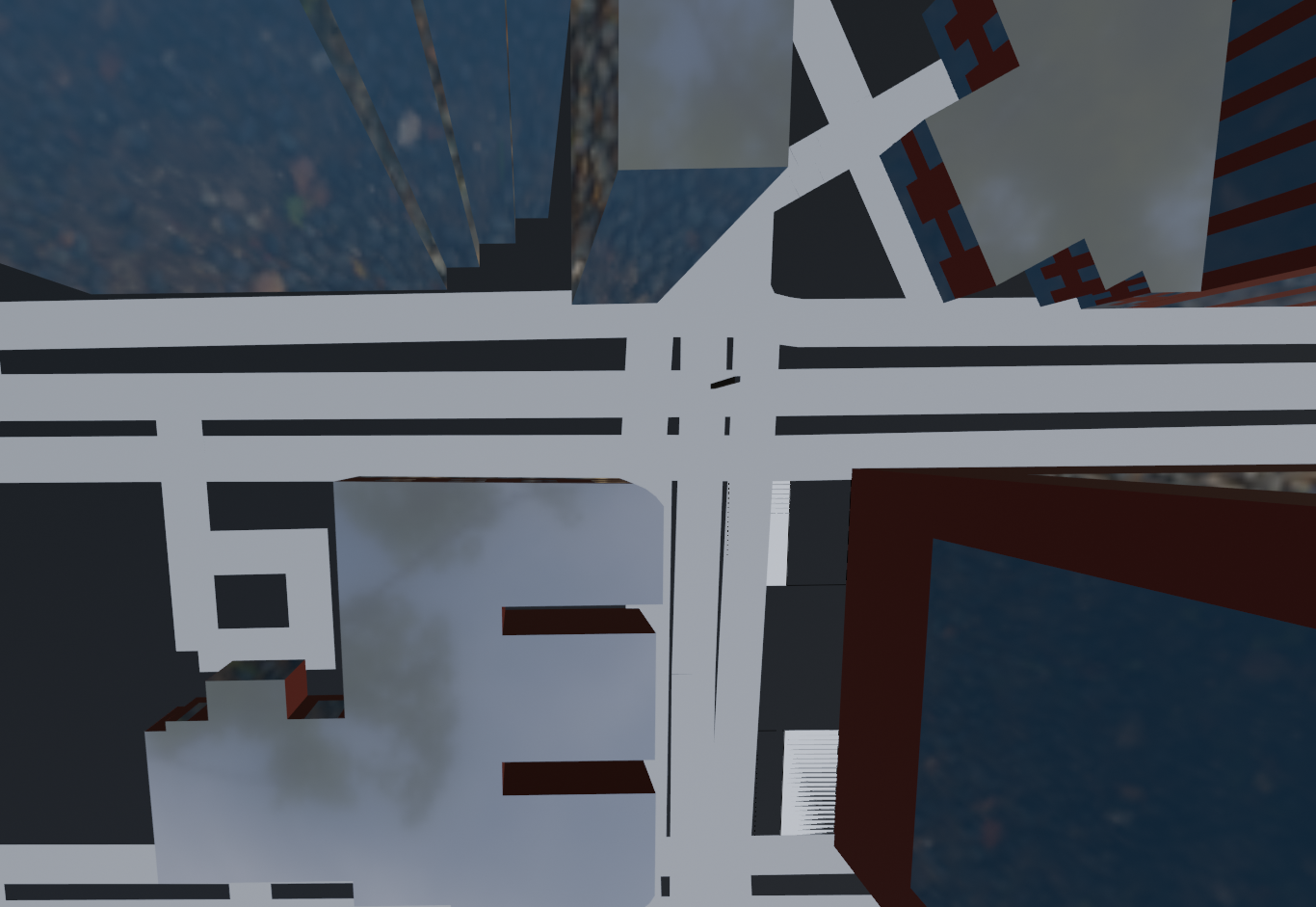}
		\caption{Blueprint Dallas Scene}
	\end{subfigure}
	\begin{subfigure}{0.4\linewidth}
        \centering
        \includegraphics[width=\linewidth]{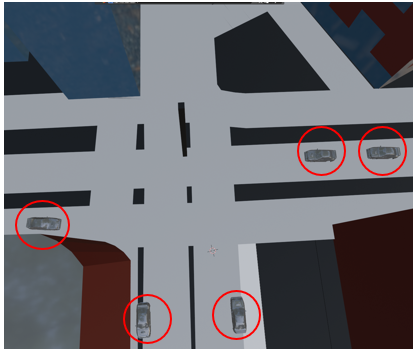}
        		\caption{Dynamic Dallas Scene}
	\end{subfigure}
        \begin{subfigure}{0.4\linewidth}
        \centering
        \includegraphics[width=\linewidth]{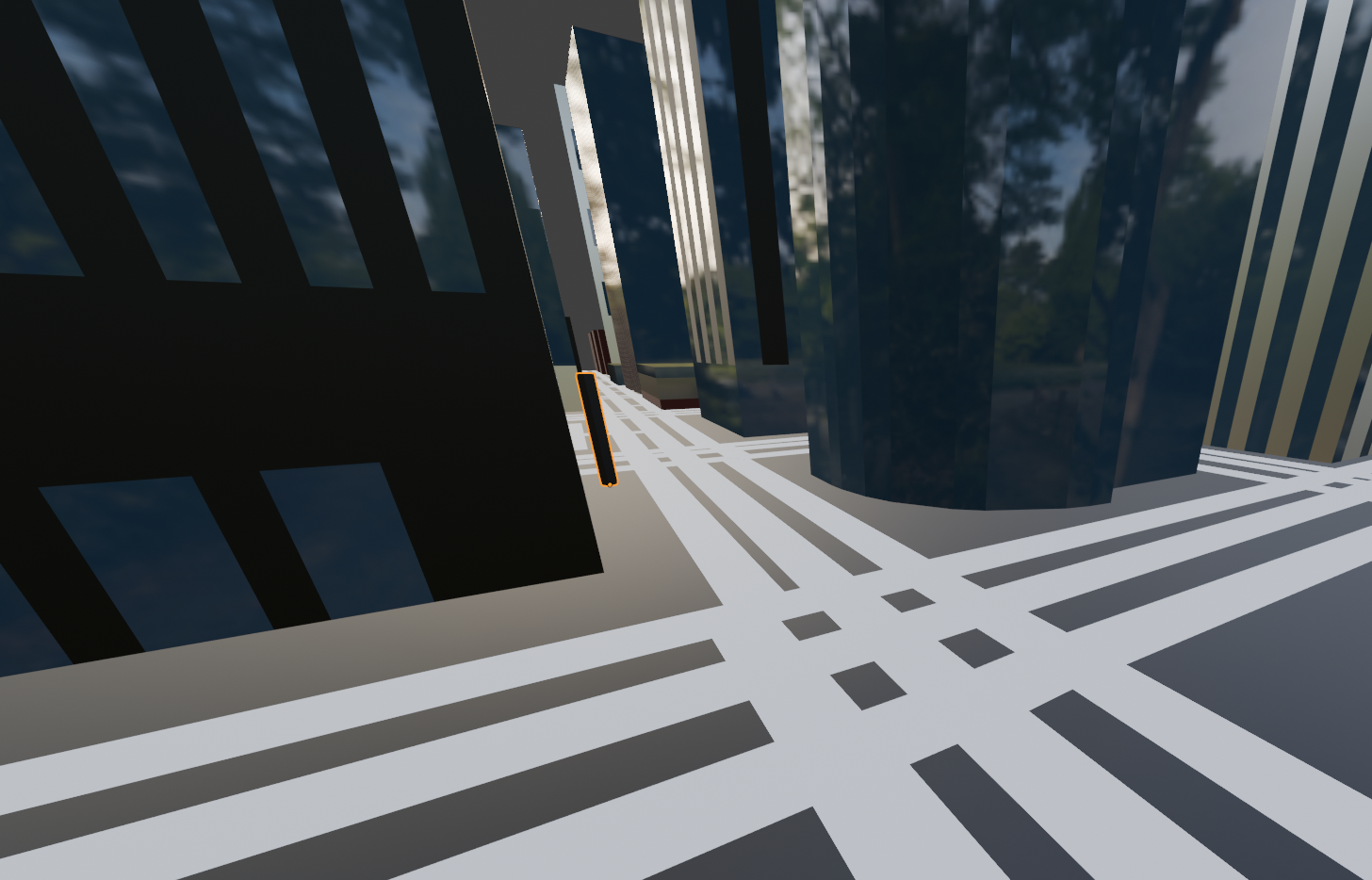}
		\caption{Blueprint  Houston Scene}
	\end{subfigure}
	\begin{subfigure}{0.4\linewidth}
        \centering
        \includegraphics[width=\linewidth]{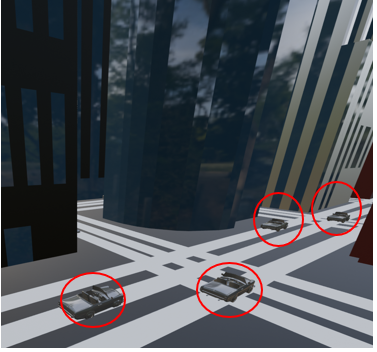}
        		\caption{Dynamic Houston Scene}
	\end{subfigure} \vspace{-0.05in}
  \caption{A snapshot of the dynamic digital twins (shown in second column) and corresponding blueprint twins (in the first column), details in Sec.~\ref{subsubsec:testbed_dynamic_twin}. We have circled the added objects in  the figure to highlight dynamicity of the scenario.}
    \vspace{-0.1in}
	\label{fig:dynamic_twins}
\end{figure}

While the blueprint digital twins provide the training data, however it does not incorporate the mobile objects of a dynamic environment. Hence, we generate another set of twins of the same scenarios by adding various mobile objects within the scene. These dynamic digital twins replicate the real-time scenarios. The details of various added dynamic objects are listed below:

\noindent $\bullet$ {\bf Lab Cubicle Dynamic Digital Twins.} We add an additional chair, adjust the orientation of three chairs, and randomly position two people in the scene, while modeling all added objects with their respective ITU-defined materials. The representation of the dynamic and corresponding blueprint twins are presented in Fig.~\ref{fig:dynamic_twins} (b) and (a), respectively.

\noindent $\bullet$ {\bf Lab Meeting Dynamic Digital Twin.} We place two laptops on the table, add two chairs and a stack of boxes, and modify the orientation of two chairs within the scene, ensuring all objects are modeled with their ITU-defined materials. The generated twins are shown in Fig.~\ref{fig:dynamic_twins} (c) and (d).

\noindent $\bullet$ {\bf Downtown Dallas Dynamic Digital Twin.} We randomly position five cars and five buses on two roads to simulate mobility within the scene, representing all objects with their ITU-defined materials. The representations are shown in Fig.~\ref{fig:dynamic_twins} (e) and (f).

\noindent $\bullet$ {\bf Downtown Houston   Dynamic Digital Twin.} We place four cars and two buses at random locations on two roads to show mobility in the scene, using ITU-specified materials to represent all objects. The generated twin and corresponding blueprint is shown in Fig.~\ref{fig:dynamic_twins} (h) and (g), respectively.


\subsection{Propagation Modeling}
\vspace{-0.02in}
\label{subsec:testbed_prop_model}
The propagation modeling process in Sionna RT commences with the integration of Sionna with the {\em Low Level Virtual Machine (LLVM) toolchain} to ensure smooth compatibility. This integration guarantees seamless scene loading, transmitters and receivers creation and configuration, replicating real-world characteristics within the scene. To accurately model RF propagation, signal behavior, and communication performance within a scene, Sionna classifies all scene components based on their {\em radio material} types and associated {\em material properties}. The transmitter and receiver antennas are configured as  $8\times2$ {\tt tr38901}~\cite{nvidia_rt_details} with dual polarization and $1\times1$ {\em diapole} planner array, respectively. 
We calculate propagation paths using the {\tt compute\_paths} function in Sionna to generate ray-traced data and produce propagation map by setting {\tt max-depth = 5} and {\tt num-samples = 200}. Overall, our Sionna based implementation ensures improved computational efficiency by incrementally updating the ray paths based on changes in the receiver (or robot) location~\cite{nvidia_rt_details}. 

\begin{figure}
	\centering
	\begin{subfigure}{0.3\linewidth}
        \centering
        \includegraphics[width=\linewidth]{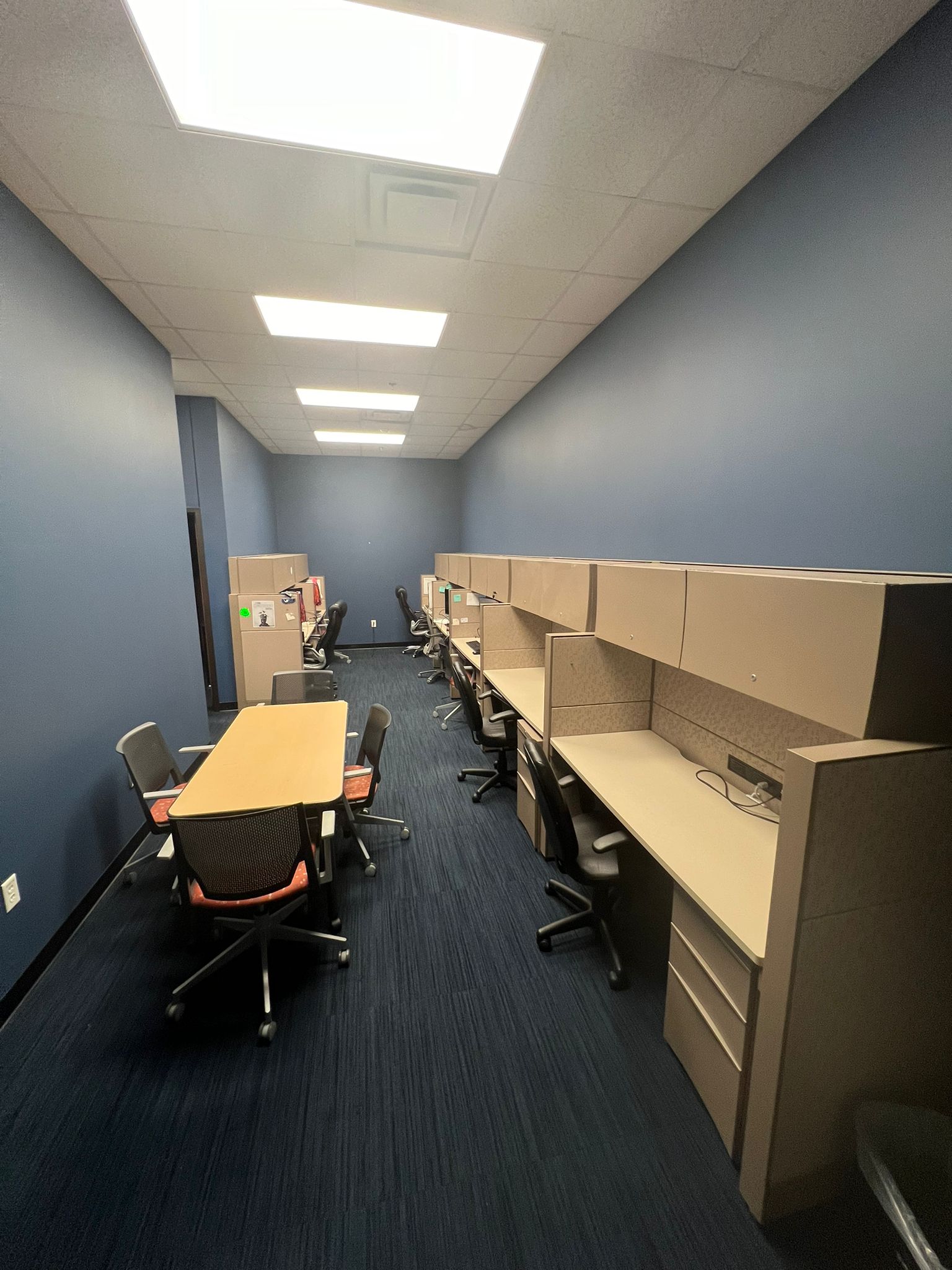}
		\caption{Lab Cubicle: RW}
	\end{subfigure}
        \begin{subfigure}{0.3\linewidth}
        \centering
        \includegraphics[trim={0cm 0 3cm 0},clip, rotate = 270, width=0.95\linewidth]{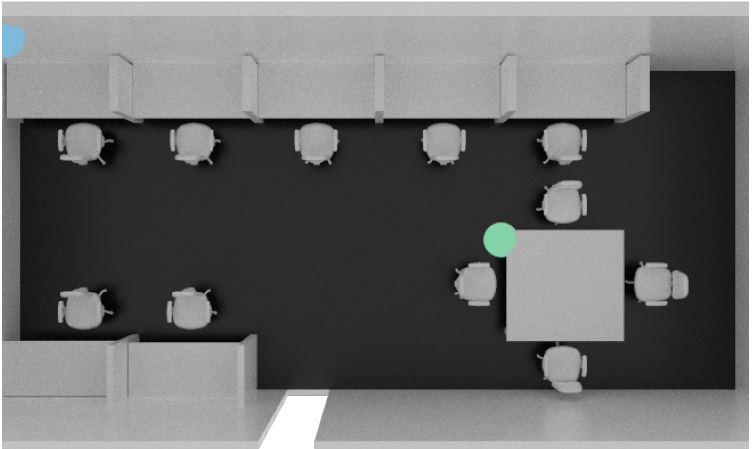}
        		\caption{Lab Cubicle: DT}
	\end{subfigure} 
	\begin{subfigure}{0.3\linewidth}
        \centering
        \includegraphics[trim={1.2cm 2cm 2.6cm 2.7cm},clip, rotate = 270, width=0.7\linewidth]{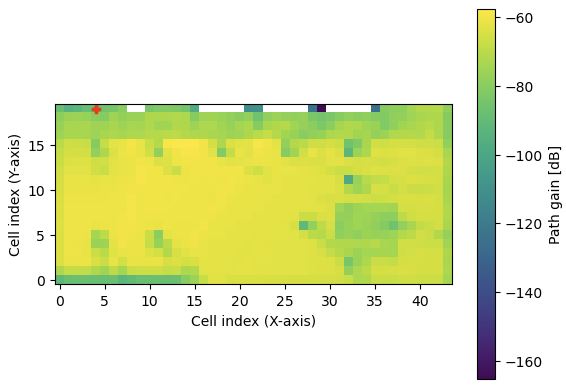}
        		\caption{Lab Cubicle: CM}
	\end{subfigure} 
 	\begin{subfigure}{0.3\linewidth}
        \centering
        \includegraphics[trim={0cm 0 3cm 0cm},clip, width=\linewidth]{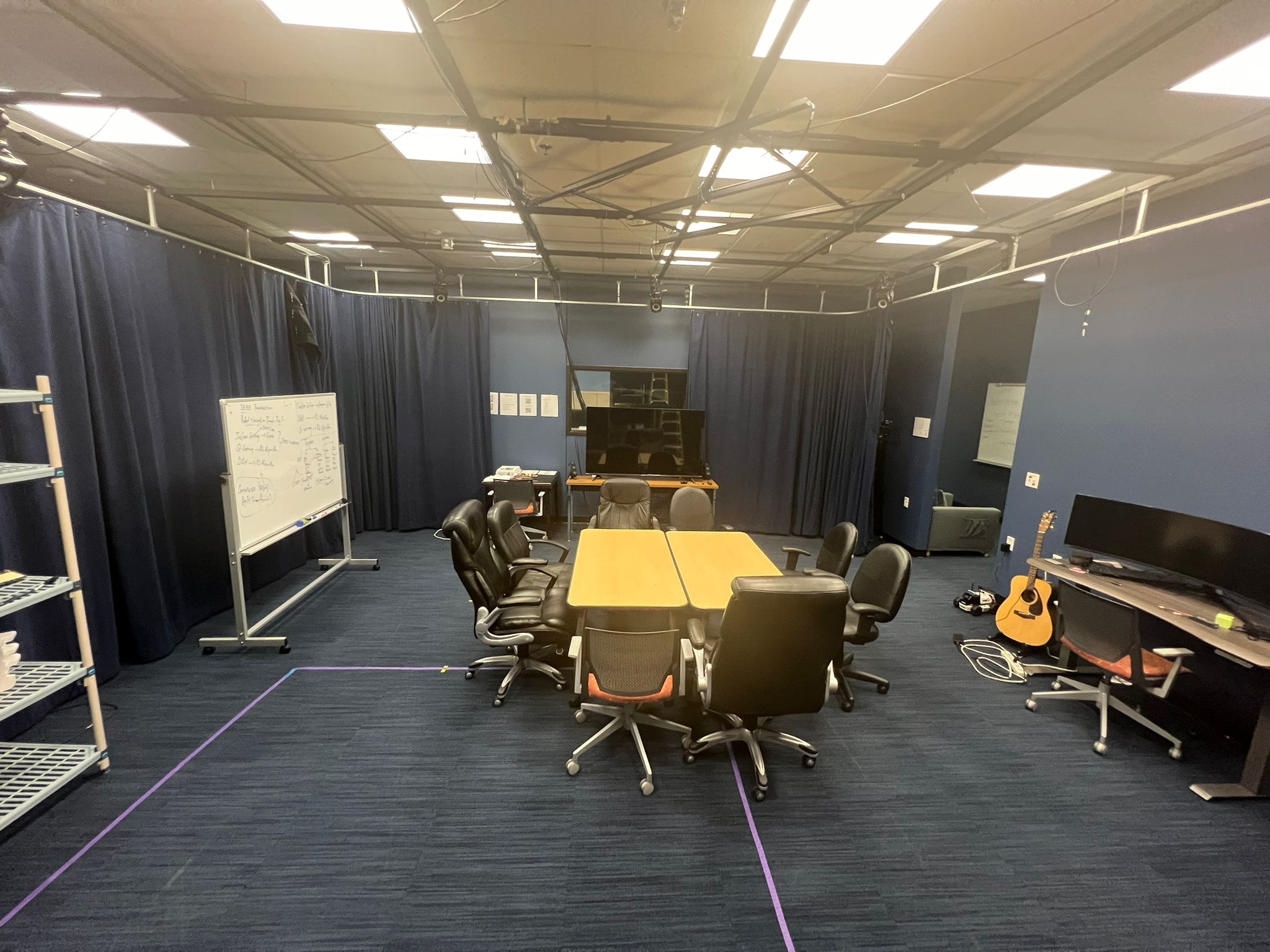}
		\caption{Lab Meeting: RW}
	\end{subfigure}
        \begin{subfigure}{0.3\linewidth}
        \centering
        \includegraphics[trim={0cm 0 0cm 0cm},clip, width=0.85\linewidth]{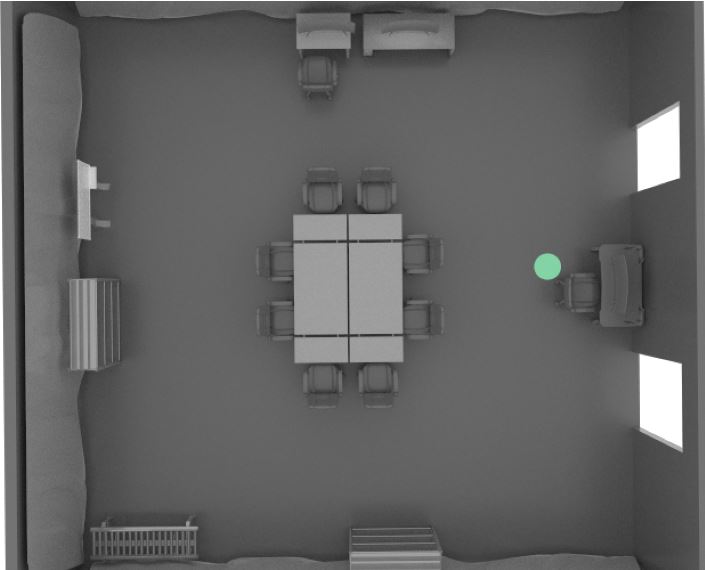}
		\caption{Lab Meeting: DT}
	\end{subfigure}
	\begin{subfigure}{0.3\linewidth}
        \centering
        \includegraphics[trim={1cm 1cm 2.2cm 0cm},clip, width=0.75\linewidth]{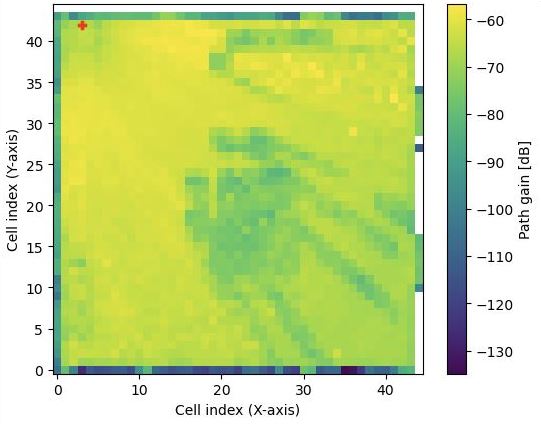}
        		\caption{Lab Meeting: CM}
	\end{subfigure} 
 	\begin{subfigure}{0.3\linewidth}
        \centering
        \includegraphics[width=0.65\linewidth]{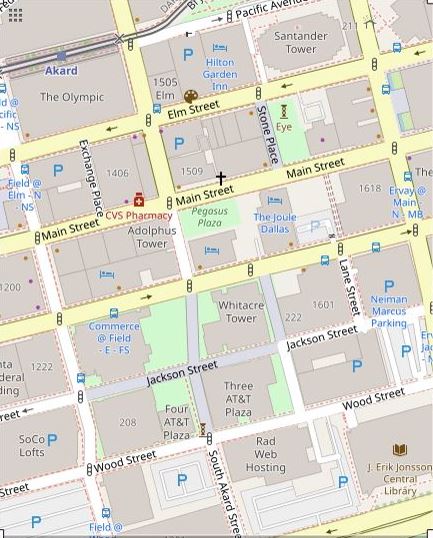}
		\caption{Dallas: RW}
	\end{subfigure}
	\begin{subfigure}{0.3\linewidth}
        \centering
        \includegraphics[trim={0cm 0 2cm 0},clip, width=0.85\linewidth]{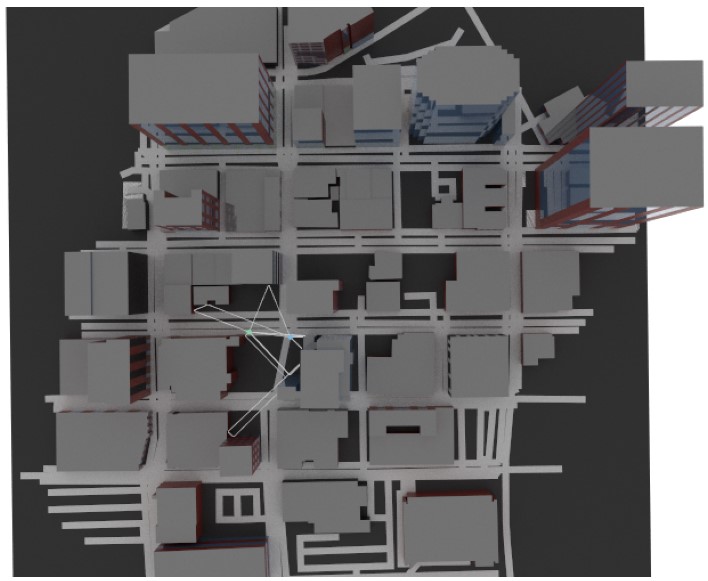}
        		\caption{Dallas: DT}
	\end{subfigure}
        \begin{subfigure}{0.3\linewidth}
        \centering
        \includegraphics[trim={0cm 0 2cm 0cm},clip, width=\linewidth]{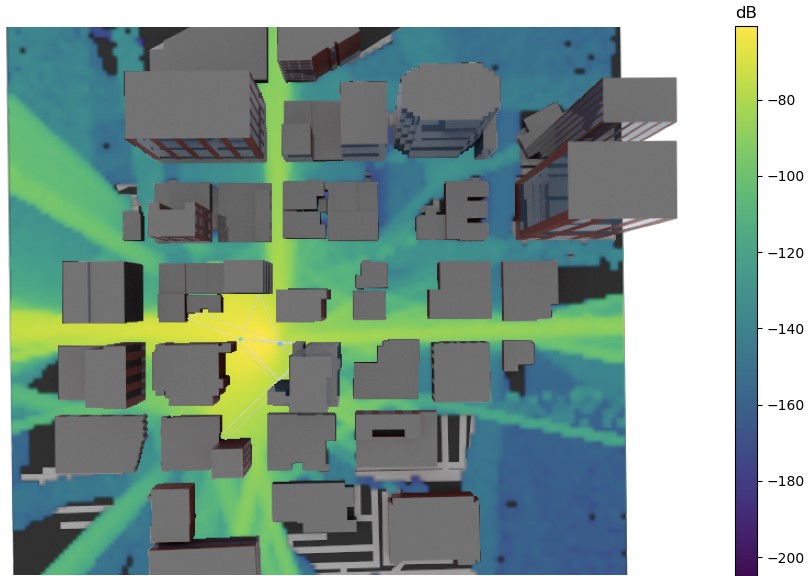}
        		\caption{Dallas: CM}
	\end{subfigure}
        \begin{subfigure}{0.3\linewidth}
        \centering
        \includegraphics[width=0.7\linewidth]{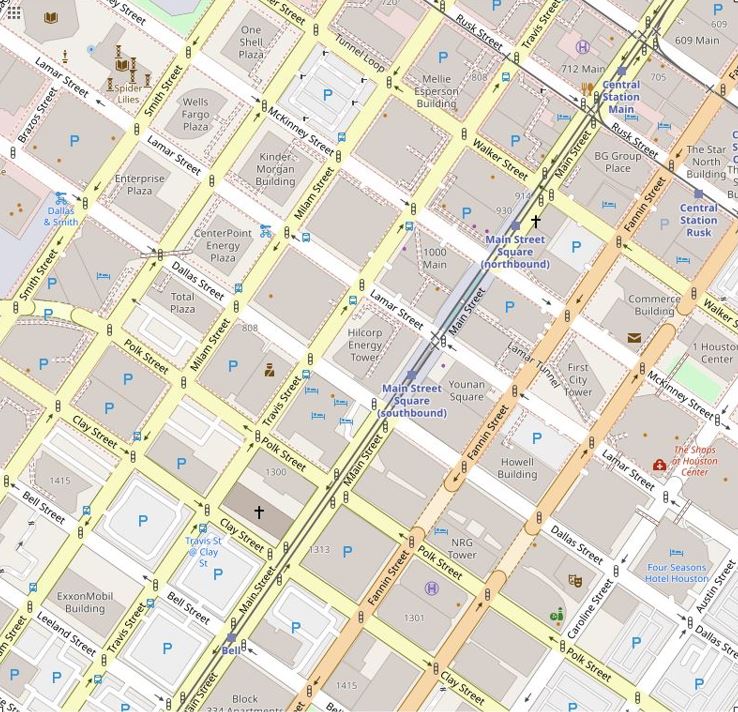}
		\caption{Houston: RW}
	\end{subfigure}
        \begin{subfigure}{0.3\linewidth}
        \centering
        \includegraphics[trim={0cm 0 2cm 0},clip, width=0.8\linewidth]{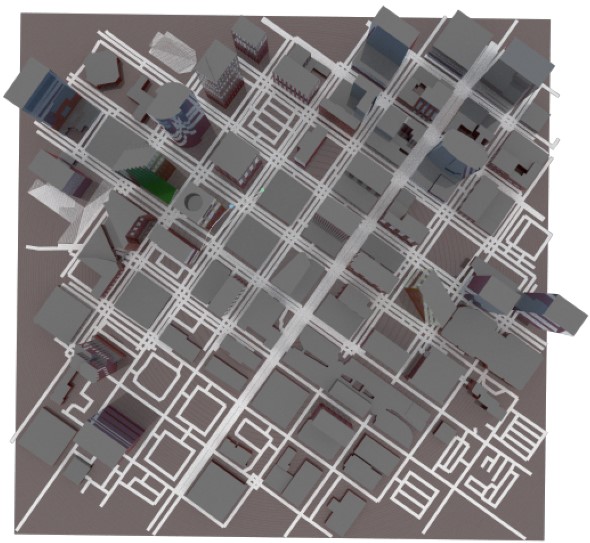}
		\caption{Houston: DT}
	\end{subfigure}
        \begin{subfigure}{0.3\linewidth}
        \centering
        \includegraphics[trim={0cm 0 2cm 0},clip, width=\linewidth]{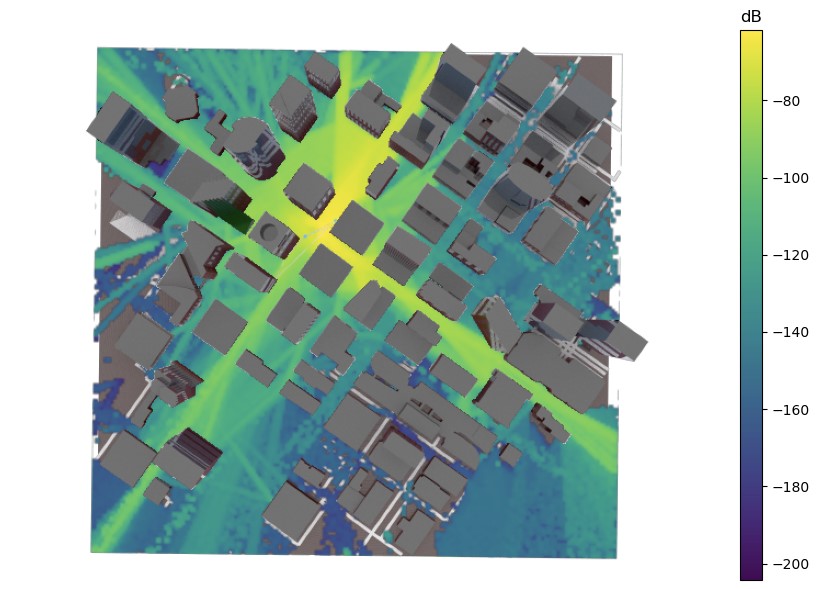}
		\caption{Houston: CM}
	\end{subfigure}\vspace{-0.05in}
  \caption{The scene maps and coverage maps for both indoor and outdoor environments (details in Sec.~\ref{Sec:proposed_mtwo}). RW, CM and DT represents real-world, coverage map and digital twin, respectively.}
  \vspace{-0.1in}
	\label{fig:blueprint_twin}
\end{figure}

\begin{figure}
	\centering
	\begin{subfigure}{0.4\linewidth}
        \centering
        \includegraphics[width=\linewidth]{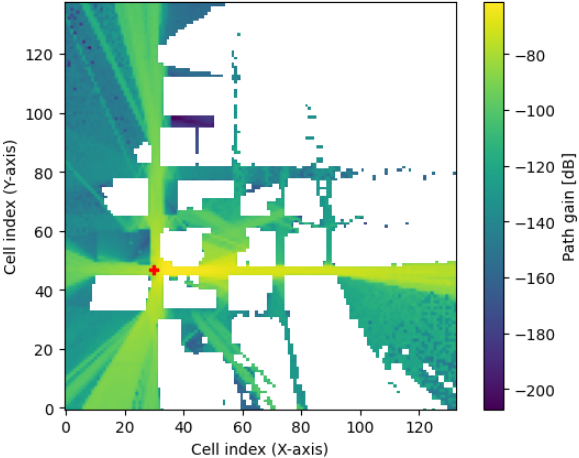}
		\caption{Dallas Q1 CM}
	\end{subfigure}
	\begin{subfigure}{0.4\linewidth}
        \centering
        \includegraphics[width=\linewidth]{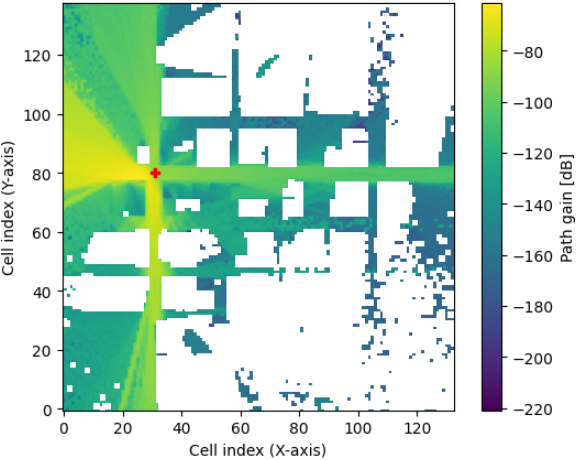}
        		\caption{Dallas Q2 CM}
	\end{subfigure} 
 	\begin{subfigure}{0.4\linewidth}
        \centering
        \includegraphics[width=\linewidth]{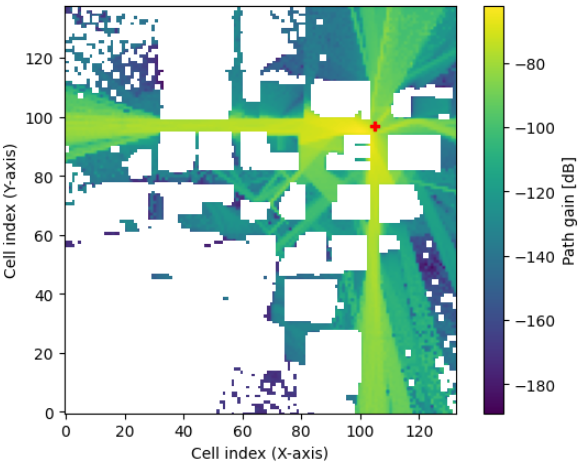}
		\caption{Dallas Q3 CM}
	\end{subfigure}
	\begin{subfigure}{0.4\linewidth}
        \centering
        \includegraphics[width=\linewidth]{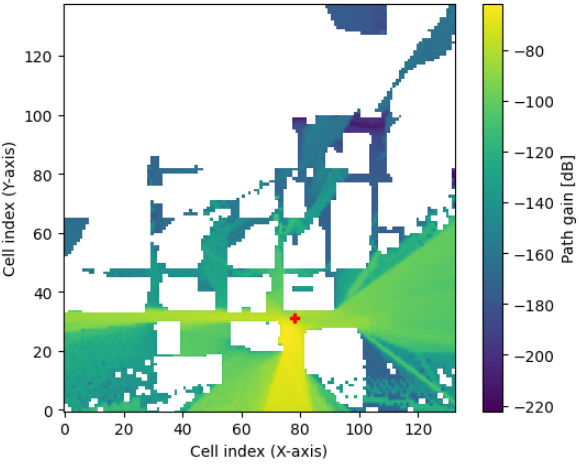}
        		\caption{Dallas Q4 CM}
	\end{subfigure} 
 	\begin{subfigure}{0.4\linewidth}
        \centering
        \includegraphics[width=\linewidth]{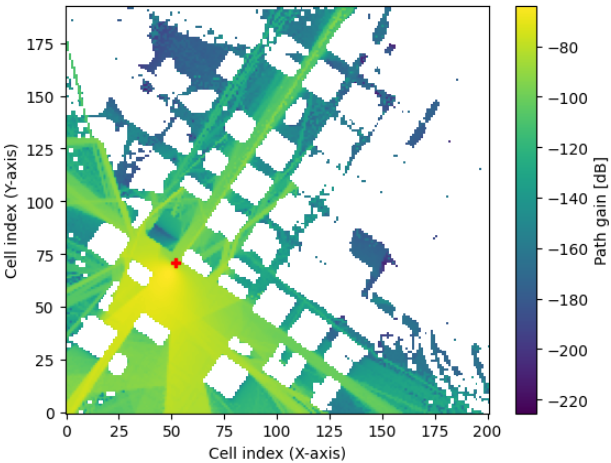}
		\caption{Houston Q1 CM}
	\end{subfigure}
	\begin{subfigure}{0.4\linewidth}
        \centering
        \includegraphics[width=\linewidth]{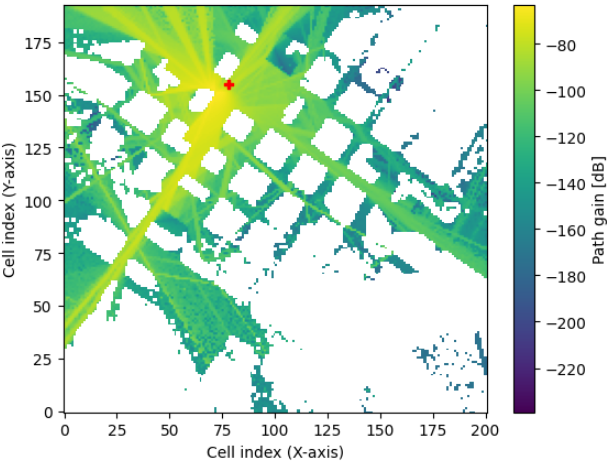}
        		\caption{Houston Q2 CM}
	\end{subfigure}
        \begin{subfigure}{0.4\linewidth}
        \centering
        \includegraphics[width=\linewidth]{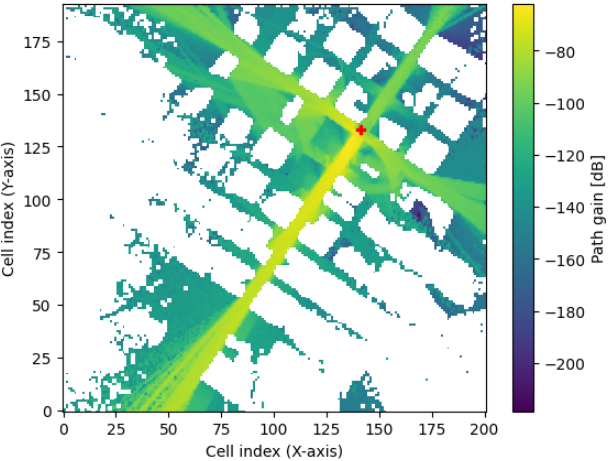}
		\caption{Houston Q3 CM}
	\end{subfigure}
	\begin{subfigure}{0.4\linewidth}
        \centering
        \includegraphics[width=\linewidth]{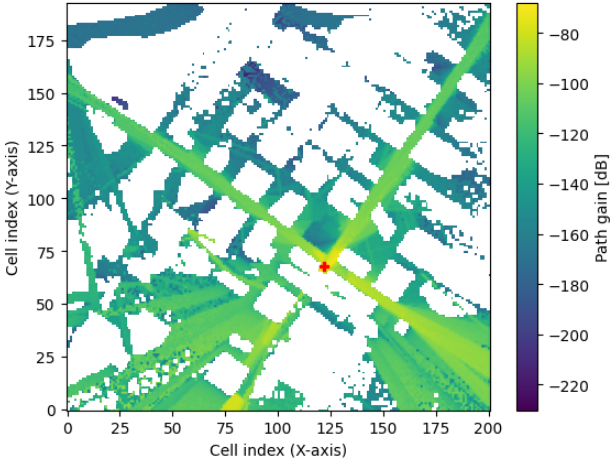}
        		\caption{Houston Q4 CM}
	\end{subfigure} \vspace{-0.1in}
  \caption{The coverage maps for the four quadrants for the outdoor Dallas and Houston blueprint digital twin scenarios (Details in Sec.~\ref{subsubsec:testbed_blueprint_twin}).}\vspace{-0.05in}
	\label{fig:quadrant_CM}
\end{figure}

Overall, the coverage maps corresponding to the propagation models of the blueprint digital twins are presented in Fig.~\ref{fig:blueprint_twin} (c, f, i, l) and Fig.~\ref{fig:quadrant_CM} for quadrant-wise experiments. In these plots, the obstacles in the physical world yields to lower signal strength regions. The white regions of Fig.~\ref{fig:quadrant_CM} represents the various buildings, hence absence of any signals.

\subsection{Generated Datasets}
\vspace{-0.03in}
\label{subsec:testbed_generated_dataset}
Following propagation modeling, a ray-tracing dataset is extracted comprising channel coefficients $\Theta$, phase shifts $\phi$, azimuth angles of departure $\alpha_d$ and arrival $\alpha_a$, as well as zenith angles of departure $\zeta_d$ and arrival $\zeta_a$ for each cell of the generated propagation map. Overall we have generated the ray-tracing datasets corresponding to: (a) all four blueprint digital twins and (b) all four dynamic digital twins.

\noindent {\bf (a) Blueprint Datasets. }
At first, we generate the ray-tracing data for each of the four \textcolor{black}{static} blueprint digital twins. The dataset encompasses 4692 rows, equating to 4692 states, with a total size of 632KB. Various statistics of the dataset are presented in Table.~\ref{tab:blueprint_dataset}.

For collecting the ray-tracing data for Dallas and Houston downtown, we divide the total downtown region in four quadrants, following the realistic scenario of having multiple base-stations or access points for a city downtown area. The details about the transmitter and receiver positions specific to those are presented in Tables~\ref{tab:dallas_dataset} and \ref{tab:houston_dataset}. The third and fourth rows of Table.~\ref{tab:blueprint_dataset}, shows the statistics of  the merged data for all quadrants. These datasets will be released for public usage upon acceptance of this article.


\begin{table}[!t]
\centering
 \caption{The description of the $\name$ dataset encompassing the blueprint digital twins.}
\resizebox{0.9\linewidth}{!}{
\begin{tabular}{|c|c|c|c|}
\hline
\textbf{Scene Name}  & \textbf{Grid Size} &   \textbf{Data Size} & \textbf{Data Row}\\
\hline
\hline
Cubicle & [42, 18] & 95 KB & 757\\
\hline
Meeting Room & [38, 40] & 170 KB & 1521\\
\hline
Downtown Dallas & [137, 130] & 39 MB  & 456483\\
\hline
Downtown Houston & [507, 473] & 20 MB & 232363\\
\hline
\end{tabular}}\vspace{-0.1in}
 \label{tab:blueprint_dataset}
\end{table}


\begin{table}[!t]
\centering
 \caption{Overview of the scene configuration in downtown Dallas.}\vspace{-0.07in}
\resizebox{0.9\linewidth}{!}{
\begin{tabular}{|c|c|c|}
\hline
\textbf{Downtown Dallas}  & \textbf{Transmitter Position} &   \textbf{Receiver Position}\\
\hline
\hline
1st Quadrant (Q1) & [-170, -92, 18] & [-80, -92, 1.5] \\
\hline
2nd Quadrant (Q2) & [-165, 74, 18] & [-319, 0, 1.5]\\
\hline
3rd Quadrant (Q3) & [205, 159, 18] & [0, 0, 1.5]\\
\hline
4th Quadrant (Q4) & [70, -170, 18] & [0, -328, 1.5]\\
\hline
\end{tabular}}\vspace{-0.08in}
 \label{tab:dallas_dataset}
\end{table}

\begin{table}[!t]
\centering
 \caption{Overview of the scene configuration in downtown Houston.}\vspace{-0.05in}
\resizebox{0.9\linewidth}{!}{
\begin{tabular}{|c|c|c|}
\hline
\textbf{Downtown Houston}  & \textbf{Transmitter Position} &   \textbf{Receiver Position}\\
\hline
\hline
1st Quadrant (Q1) & [-258, -118, 20] & [-507, -473, 1.5] \\
\hline
2nd Quadrant (Q2) & [-127, 304, 20] & [-507, 0, 1.5]\\
\hline
3rd Quadrant (Q3) & [189, 195, 20] & [0, 0, 1.5]\\
\hline
4th Quadrant (Q4) & [96, -134, 20] & [0, -473, 1,5]\\
\hline
\end{tabular}}
\vspace{-0.2in}
 \label{tab:houston_dataset}
\end{table}

\noindent {\bf (b) Dynamic Datasets. } The dynamic datasets are collected by running the ray-tracing algorithm on the dynamic twins presented in Sec.~\ref{subsubsec:testbed_dynamic_twin}. The Dallas dynamic twin is created on by updating the Dallas quadrant 3 (Q3) blueprint twin with the added objects mentioned in Sec.~\ref{subsubsec:testbed_dynamic_twin}. Similarly, the Houston dynamic twin is created on top of the Houston quadrant 2 (Q2) blueprint twin. The statistics of the generated ray-tracing data for the dynamic twins are shown in Table~\ref{tab:dynamic_dataset}.

\begin{table}[!t]
\centering
 \caption{The description of the $\name$ dataset encompassing the dynamic digital twins.}
\resizebox{0.9\linewidth}{!}{
\begin{tabular}{|c|c|c|c|}
\hline
\textbf{Scene Name}  & \textbf{Grid Size} &   \textbf{Data Size} & \textbf{Data Row}\\
\hline
\hline
Cubicle & [42, 18] & 90 KB & 757\\
\hline
Meeting Room & [38, 40] & 142 KB & 1521\\
\hline
Downtown Dallas & [137, 130] & 10 MB  & 123825\\
\hline
Downtown Houston & [180, 200] & 5 MB & 61216\\
\hline
\end{tabular}}\vspace{-0.2in}
 \label{tab:dynamic_dataset}
\end{table}

\begin{remark}
The overall generated ray-traced data, representing the blueprint and dynamic digital twins, are $<100MB$ and $<10MB$, which enables the DQN agent to have a faster training and inference, respectively (validates Contribution 4).
\vspace{-0.05in} 
\end{remark}


\section{Performance Evaluation}
\label{sec:performance}
\noindent
{\bf Experimental Platform.} We implement and validate the $\name$ DQN agent {\em (Module 3)} on the $\name$ dataset. The experiments are performed on an Intel (R) Xeon w7-2495x processor using Python with Pytorch, numpy, and matplotlib libraries. 

\noindent
{\bf Evaluation Metrics:} To measure the performance of the training of the robot navigation, we evaluate the total rewards at each episode. We also consider the minimum number of steps the robot takes to reach the final destination. Finally, we also present the training time and end-to-end decision delay for the $\name$ DQN agent.\vspace{-0.05in} 

\subsection{Experimental Dataset}\vspace{-0.05in}
We validate our proposed $\name$ framework using the blueprint and dynamic digital twins for two indoor and two outdoor scenarios, details in Sec.~\ref{sec:testbed}. The experiments have been configured based on various training and inference scenarios involving the blueprint and dynamic digital twins.\vspace{-0.05in}

\subsection{DQN Training}
\begin{figure} [t]
        \centering
        \includegraphics[width=0.5\linewidth]{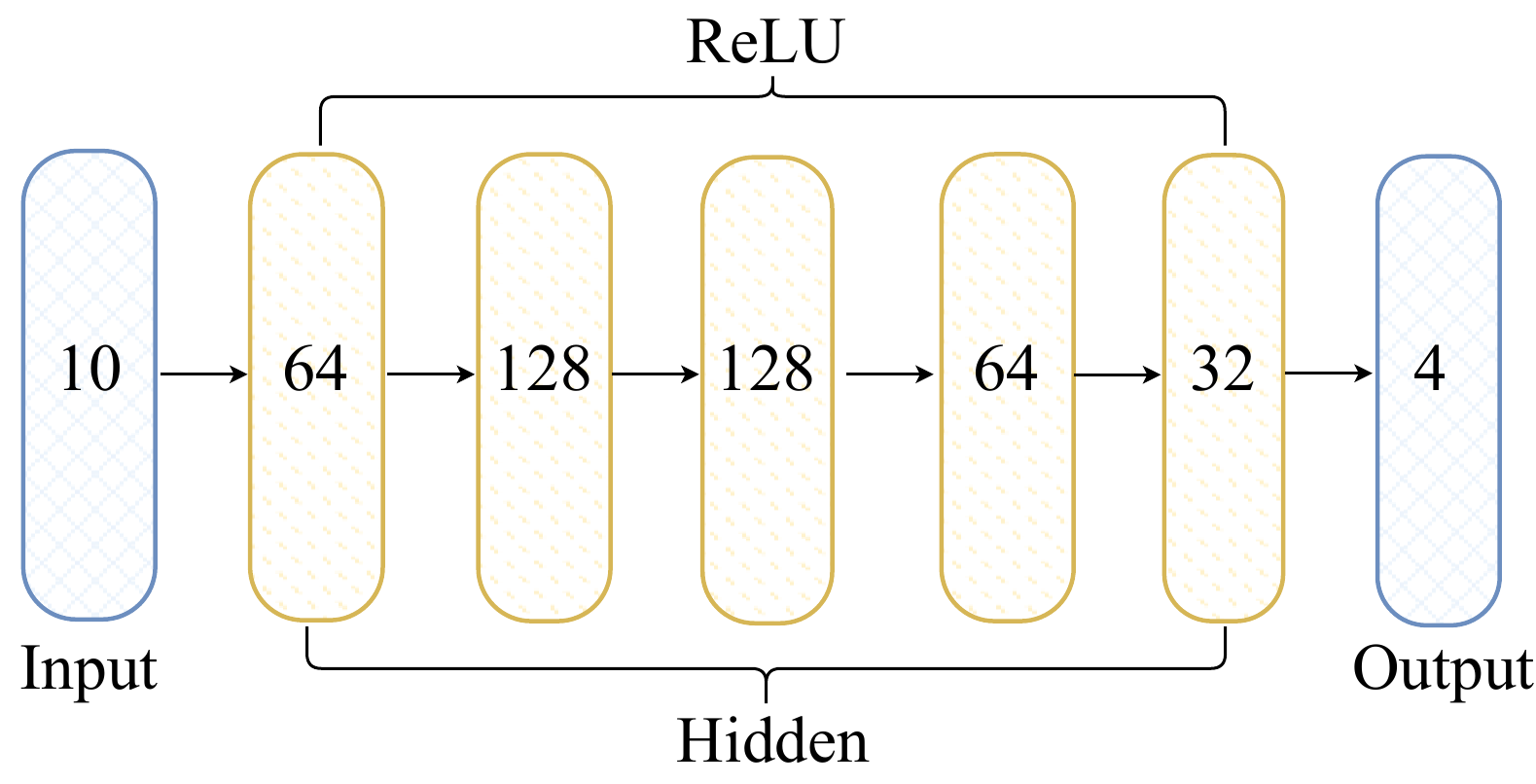}
        \vspace{-0.03in}
        \caption{Neural network architecture used in $\name$ framework.}
        \vspace{-0.03in}
        \label{fig:NN-Architecture}
        \vspace*{-10pt}
    \end{figure}

\vspace{-0.05in}
The deep neural network model we use to train our DQN agent within the $\name$ framework consists of an input layer with 10 neurons representing the 10 features that the agent received in each state which are X coordinate ($R_x^{t}$), Y coordinate ($R_y^{t}$), azimuth angle of arrival ($\alpha_a^t$), zenith angle of arrival ($\zeta_a^t$), azimuth angle of departure ($\alpha_d^t$), zenith angle of departure ($\zeta_d^t$), channel coefficient's real and imaginary values ($|\Theta|^t$), and propagation delay ($\phi^t$) within the digital twin. The details about these inputs are given in Sec.~\ref{subsubsec:state}. It is followed by five dense layers each having the rectified linear unit (ReLU) as the activation function, as shown in Fig.~\ref{fig:NN-Architecture}. The output layer had 4 {\em Dense} units with linear activation function representing the Q-values of the four actions that the agent chooses from. We trained the agent using the DQN algorithm for $80$ iterations in the outdoor environment and $500$ iterations in the indoor environment. For each iteration, we take the number of steps as long as the agent reaches the target location from the starting location. We randomly set the starting location and the target location on the map. The behavior policy $\theta$ is trained by updating the behavior DQN  architecture using the Q values from the target DQN architecture. We follow an $\epsilon$-greedy policy to learn the Q values with an exploration rate $\epsilon$ of 0.1. We use Bellman's optimality equation~\cite{10.5555/3312046} to update the Q-values. Q-values of the behaviour network is trained with reference to the Q-values of the next state from the target network discounted with a factor of $0.9$, of the action followed to reach that next state and added to the reward received in that next state. 

The source and target coordinates during training and inference of all experiments are taken randomly, considering that the coordinate is not an obstacle or building in the corresponding scenario.\vspace{-0.05in} 



\subsection{Performance of $\name$ DQN Agent (Trained/Inference on Blueprint Twin)}\vspace{-0.05in}
In this first set of experiments, we validate the $\name$ DQN agent by training and inferencing on the same scenarios.

\begin{figure}[t!]
     \centering    
     \includegraphics[width=1\linewidth]{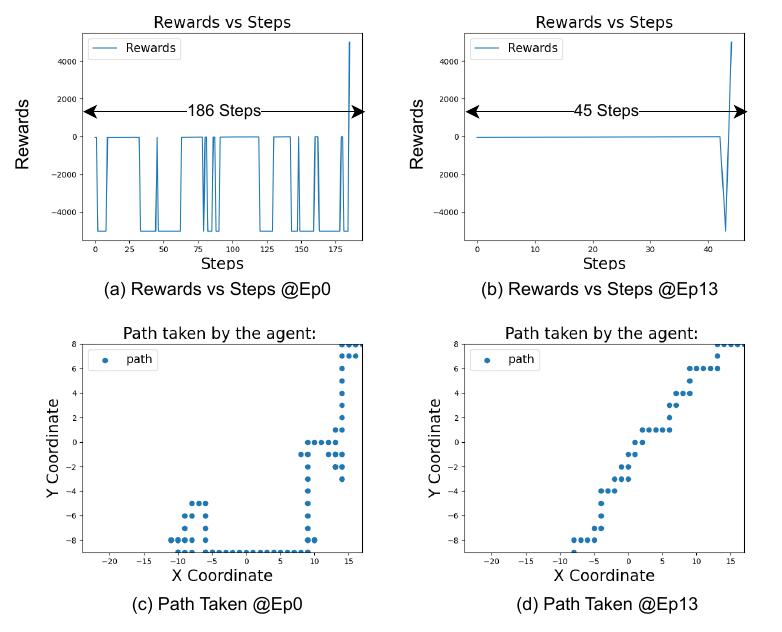}\vspace{-0.1in}
     \caption{The training performance of the $\name$ DQN agent on the blueprint twin of lab cubicle. We see 75.8\% of reduction is required number of steps to 45th episode.}
     \vspace{-0.2in}
     \label{fig:lab-cubicle-training-plot}
 \end{figure}

\begin{figure}
	\centering
	\begin{subfigure}{0.4\linewidth}
        \centering
        \includegraphics[width=\linewidth]{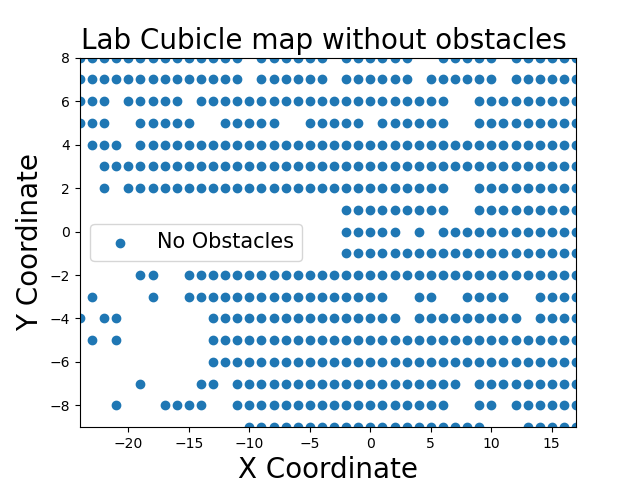}
		\caption{Cubicle}
	\end{subfigure}
	\begin{subfigure}{0.4\linewidth}
        \centering
        \includegraphics[width=\linewidth]{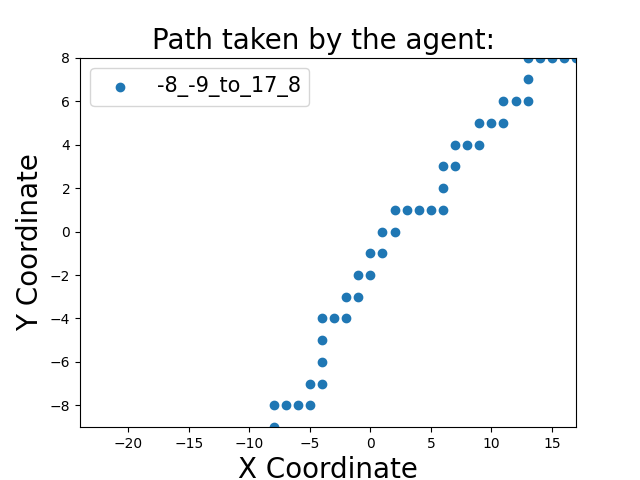}
        		\caption{Nav Path} 
	\end{subfigure} 
 	\begin{subfigure}{0.4\linewidth}
        \centering
        \includegraphics[width=\linewidth]{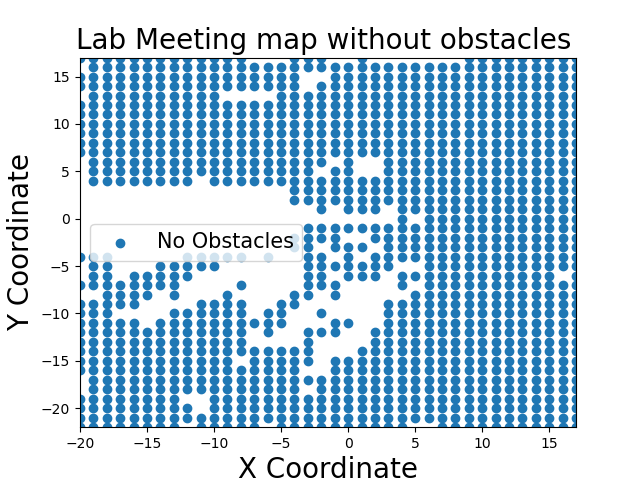}
		\caption{Meeting}
	\end{subfigure}
        \begin{subfigure}{0.4\linewidth}
        \centering
        \includegraphics[width=\linewidth]{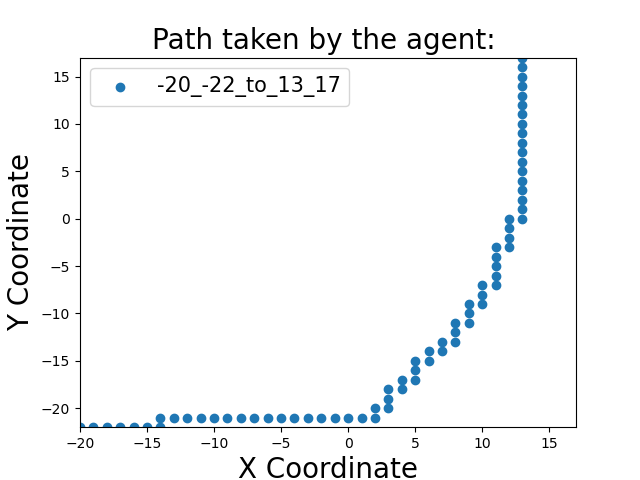}
		\caption{Nav Path} 
	\end{subfigure}\vspace{-0.05in}
     \caption{Inference on blueprint digital twins (lab cubicle and meeting space). In (a) and (c), the areas with and without collisions are illustrated for the cubicle and meeting scenes, respectively. In contrast, (b) and (d) depict the robot's successful navigation, where it follows an optimal path while avoiding obstacles, from the starting point to the target in both the cubicle and meeting scenes, respectively.}
    \vspace{-0.2in}
	\label{fig:Cubicle-Meeting Inference}
\end{figure}

\noindent $\bullet$ {\bf Lab Cubicle (Training).} For the lab cubicle, we observe that the DQN agent took more than 185 steps to reach a specific target, leading to more fluctuating reward plots in the first episode, which is episode 0 in our experiments, as shown in Fig.~\ref{fig:lab-cubicle-training-plot} (a). The larger number of dips in the reward plot signify a huge number of collisions in this plot, which later improves over training in Fig.~\ref{fig:lab-cubicle-training-plot} (b). It is also evident from Fig.~\ref{fig:lab-cubicle-training-plot} (c) that the path taken by the agent at the first episode is not optimal or shortest, rather it is zigzagged at various locations, which is due to the lack of training of the agent and the random Q-values at the beginning of the training. After training for 45 number of episodes, we observe that the DQN agent learned to reach the target while avoiding the obstacles and yielding to lesser collisions, plotted in Fig.~\ref{fig:Cubicle-Meeting Inference} (b). The path taken by the agent at this stage is also shown in Fig.~\ref{fig:lab-cubicle-training-plot} (d). The start and end coordinates are picked randomly which does not overlap with any obstacle region.

\noindent $\bullet$ {\bf Lab Cubicle (Inference).} 
The DQN agent navigates the scene following the optimal path from the starting point at the coordinates [-8, -9] to the target point at [17, 8] in the XY coordinates system, as shown in Fig.~\ref{fig:Cubicle-Meeting Inference} (b). These specific coordinates are picked randomly which does not coincide with the obstacles. Throughout navigation, the agent successfully avoids all obstacles, resulting in no collisions.

\begin{figure}[t!]
     \centering    
     \includegraphics[width=1\linewidth]{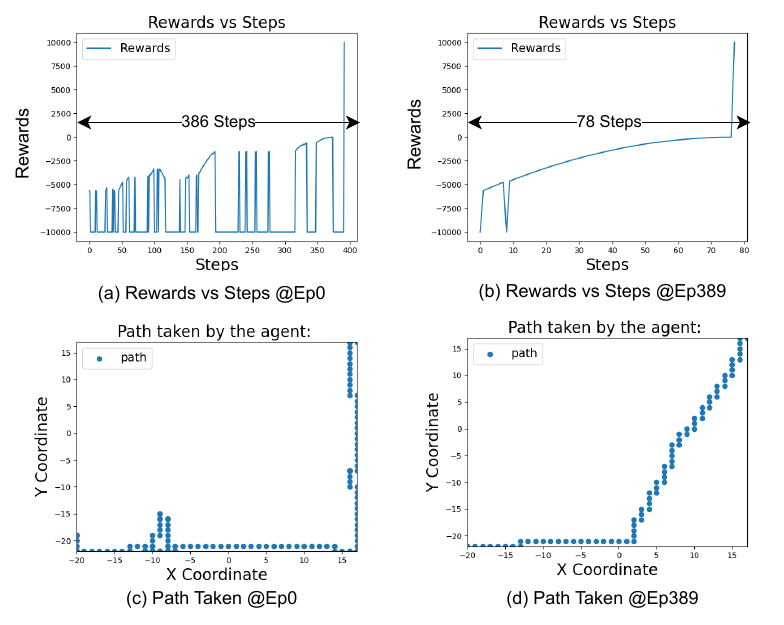}\vspace{-0.1in}
     \caption{The training performance of the $\name$ DQN agent on the blueprint twin of lab meeting space. We see 79.9\% of reduction in required number of steps at 78th episode. } 
     \vspace{-0.2in}
     \label{fig:lab-meeting-training-plot}
 \end{figure}

\noindent $\bullet$ {\bf Indoor Meeting Space (Training).} We observe similar trend of the reward values and navigation path for the meeting space scenario, the plots are shown in Fig.~\ref{fig:lab-meeting-training-plot}. 

\noindent $\bullet$ {\bf Indoor Meeting Space (Inference).} Similar to the cubicle scene, the agent navigates the meeting scene by following the optimal path from the source at coordinates [-20, 22] to the destination at coordinates [13, 17], as illustrated in Fig.~\ref{fig:Cubicle-Meeting Inference} (d). Similar to previous experiments, these specific coordinates are picked randomly which does not coincide with any obstacle. Observing the navigation area in Fig.~\ref{fig:Cubicle-Meeting Inference} (c), it is clear that the agent successfully avoids all obstacles.

\begin{figure}
     \centering    
     \includegraphics[width=1\linewidth]{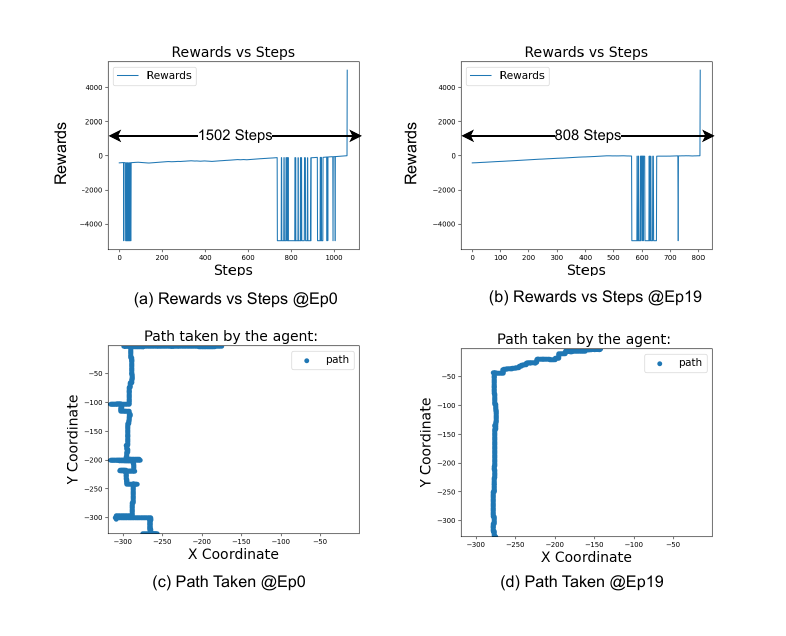}\vspace{-0.25in}
     \caption{The training performance of the $\name$ DQN agent on the blueprint twin of Dallas Q1. We see a 46.2\% of reduction in the required number of steps by the 19th episode.}
     \vspace{-0.2in}
     \label{fig:dallas-q1-training-plot}
 \end{figure}

\begin{figure}
     \centering    
     \includegraphics[width=1\linewidth]{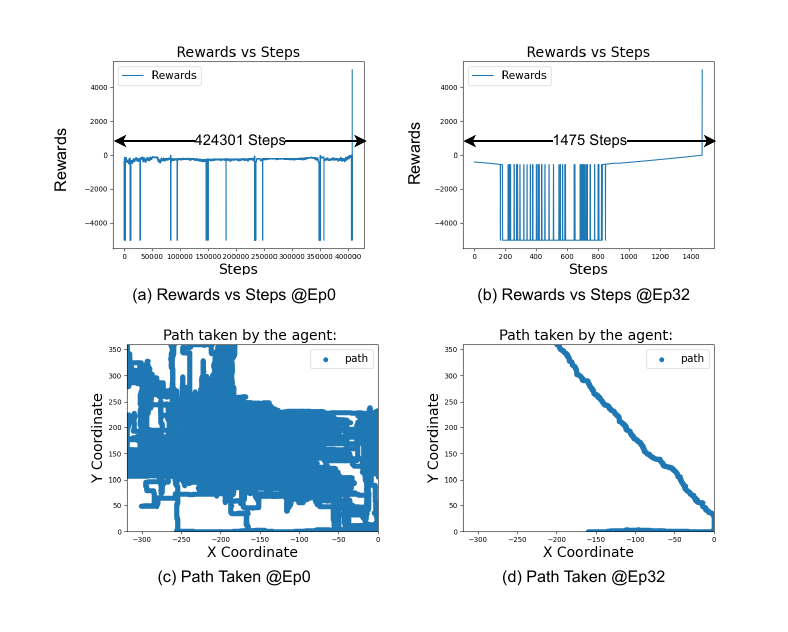}\vspace{-0.25in}
     \caption{The training performance of the $\name$ DQN agent on the blueprint twin of Dallas Q2. We see a 99.6\% of reduction in the required number of steps by the 32nd episode.}
     \vspace{-0.2in}
     \label{fig:dallas-q2-training-plot}
 \end{figure}

\begin{figure}
     \centering   
     \includegraphics[width=1\linewidth]{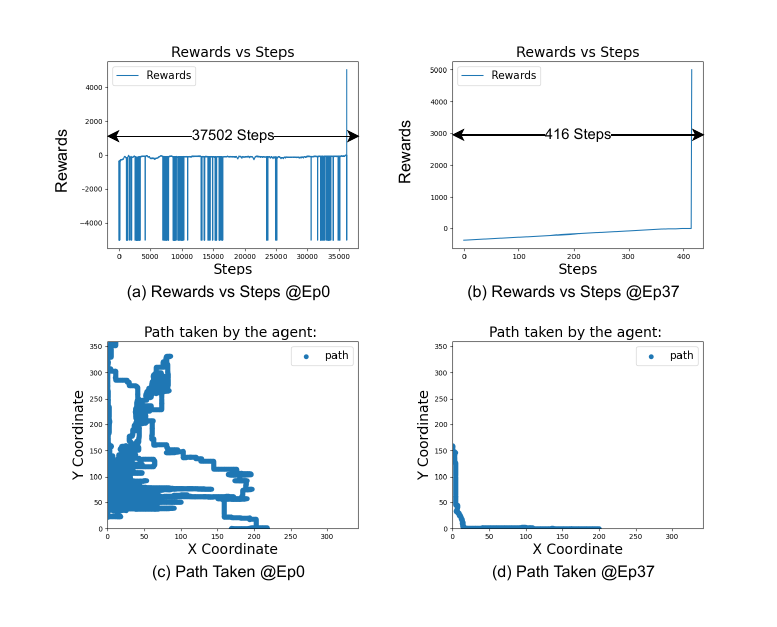}\vspace{-0.25in}
     \caption{The training performance of the $\name$ DQN agent on the blueprint twin of Dallas Q3. We see a 98.8\% of reduction in the required number of steps by the 37th episode.}
     \vspace{-0.2in}
     \label{fig:dallas-q3-training-plot}
 \end{figure}

\begin{figure}
     \centering   
     \includegraphics[width=1\linewidth]{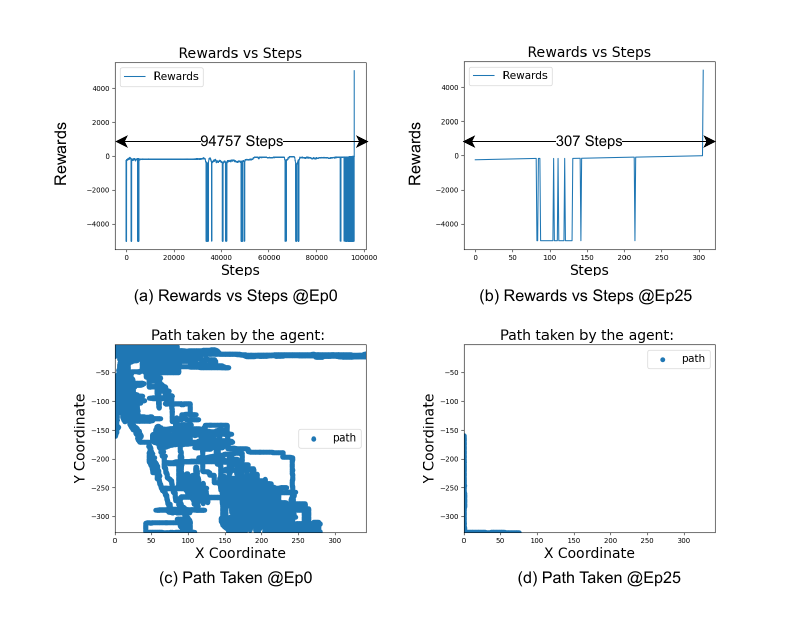}\vspace{-0.25in}
     \caption{The training performance of the $\name$ DQN agent on the blueprint twin of Dallas Q4. We see a 99.6\% of reduction in the required number of steps by the 25th episode. }
     \vspace{-0.2in}
     \label{fig:dallas-q4-training-plot}
 \end{figure}

\noindent $\bullet$ {\bf Downtown Dallas (Training).} 
In the downtown Dallas scenario, we perform quadrant-wise training over the blueprint twins, details of quadrants are in Sec.~\ref{sec:testbed}. The training performance of all four different quadrants for this case is presented in Figs.~\ref{fig:dallas-q1-training-plot}, \ref{fig:dallas-q2-training-plot}, \ref{fig:dallas-q3-training-plot}, and \ref{fig:dallas-q4-training-plot}. We observe, the DQN agent learned the optimal path around 19th, 32nd, 37th, and 25th episodes for Dallas first (Q1), second (Q2), third (Q3), and fourth (Q4) quadrants, respectively. 

\begin{figure}
	\centering
	\begin{subfigure}{0.4\linewidth}
        \centering
        \includegraphics[width=\linewidth]{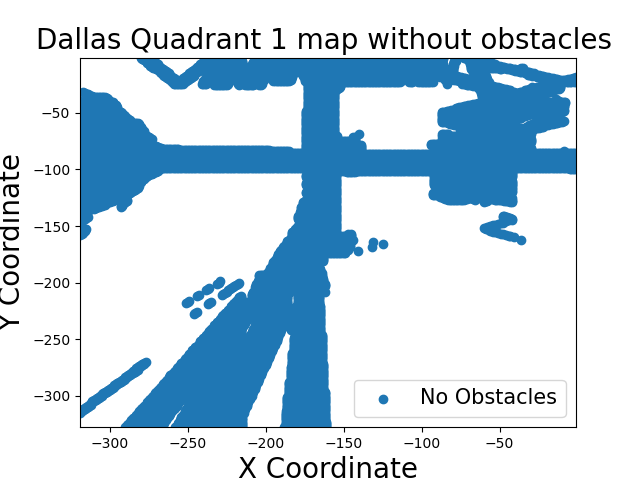}
		\caption{Dallas-Q1 Scene}
	\end{subfigure}
	\begin{subfigure}{0.4\linewidth}
        \centering
        \includegraphics[width=\linewidth]{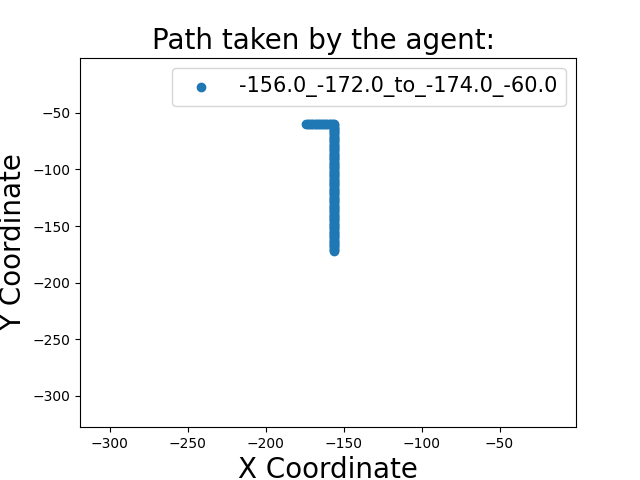}
        		\caption{Dallas-Q1 Nav Path@Infer}
	\end{subfigure}
        \begin{subfigure}{0.4\linewidth}
        \centering
        \includegraphics[width=\linewidth]{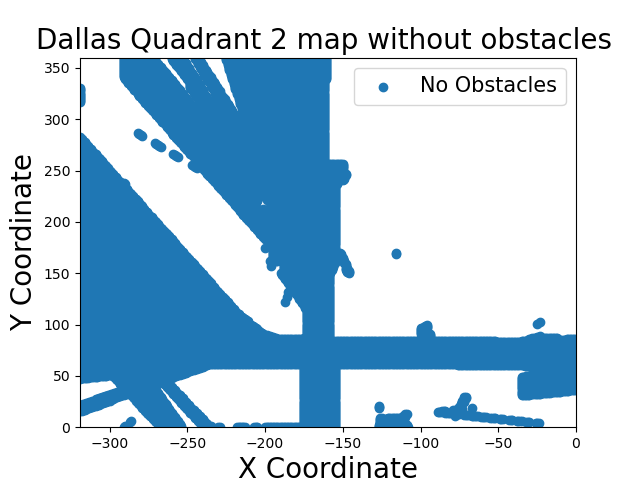}
        		\caption{Dallas-Q2 Scene}
	\end{subfigure} 
        \begin{subfigure}{0.4\linewidth}
        \centering
        \includegraphics[width=\linewidth]{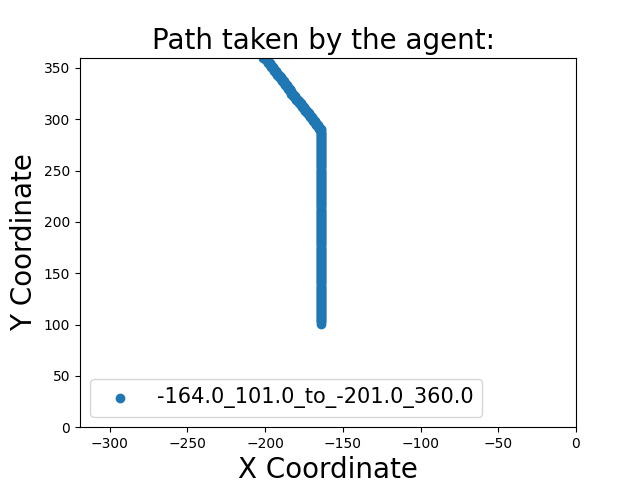}
        		\caption{Dallas-Q2 Nav Path@Infer}
	\end{subfigure} 
 	\begin{subfigure}{0.4\linewidth}
        \centering
        \includegraphics[width=\linewidth]{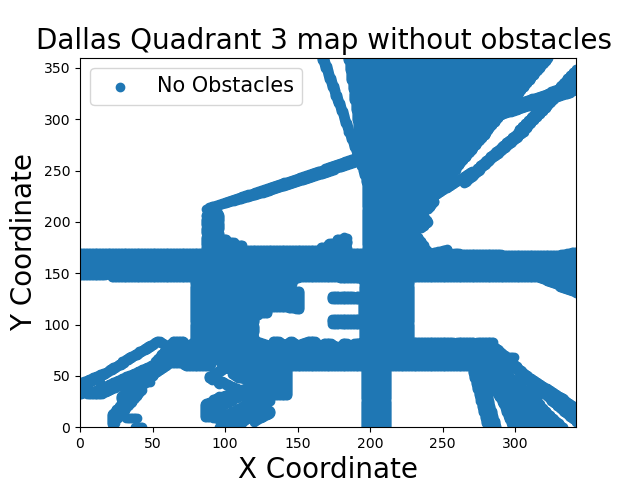}
		\caption{Dallas-Q3 Scene}
	\end{subfigure}
	\begin{subfigure}{0.4\linewidth}
        \centering
                \includegraphics[width=\linewidth]{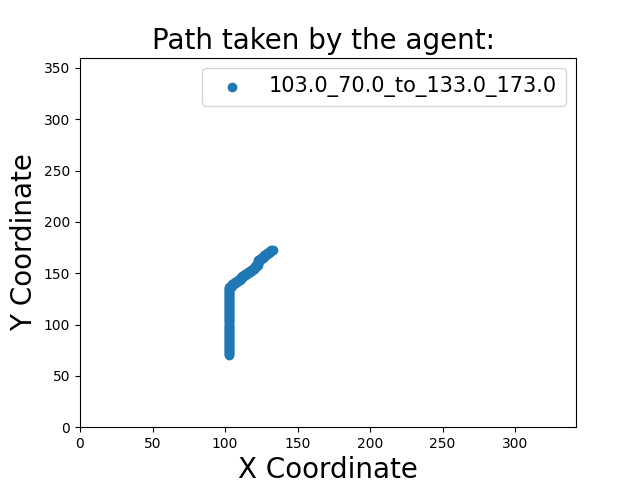}
        		\caption{Dallas-Q3 Nav Path@Infer}
	\end{subfigure} 
 	\begin{subfigure}{0.4\linewidth}
        \centering
        \includegraphics[width=\linewidth]{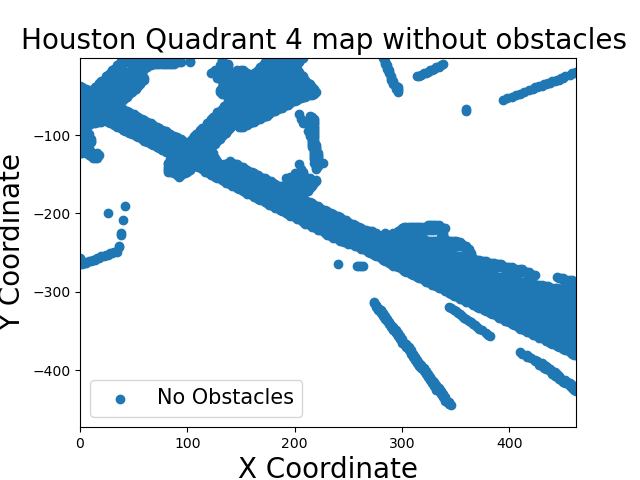}
		\caption{Dallas-Q4 Scene}
	\end{subfigure}
	\begin{subfigure}{0.4\linewidth}
        \centering
                \includegraphics[width=\linewidth]{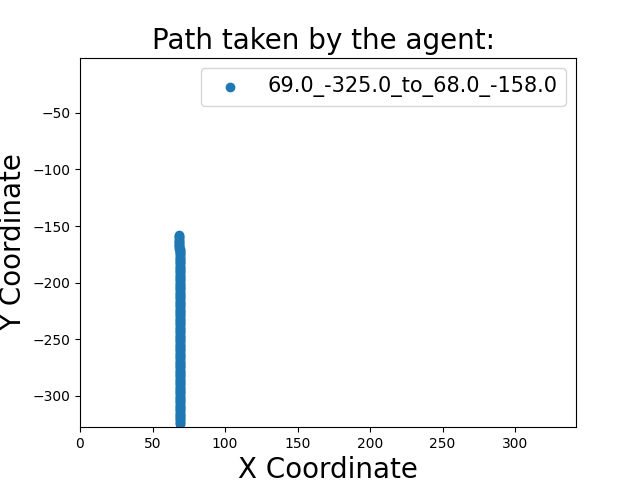}
        		\caption{Dallas-Q4 Nav Path@Infer}
	\end{subfigure}\vspace{-0.05in}
     
  \caption{Inference on blueprint digital twins (downtown Dallas). The first column (a), (c), (e), and (g), illustrates the non-obstacle regions within in each quadrant. Plots in (b), (d), (f), and (h) show the robot's successful movement from one point on the street to the target point (changing the X and Y coordinates) while avoiding obstacles in quadrants Q1, Q2, Q3, and Q4, respectively. }
    \vspace{-0.2in}
	\label{fig:Outdoors-Inference}
\end{figure}

\begin{figure}
	\centering
    \vspace{-0.3in}
	\begin{subfigure}{0.40\linewidth}
        \centering        \includegraphics[width=\linewidth]{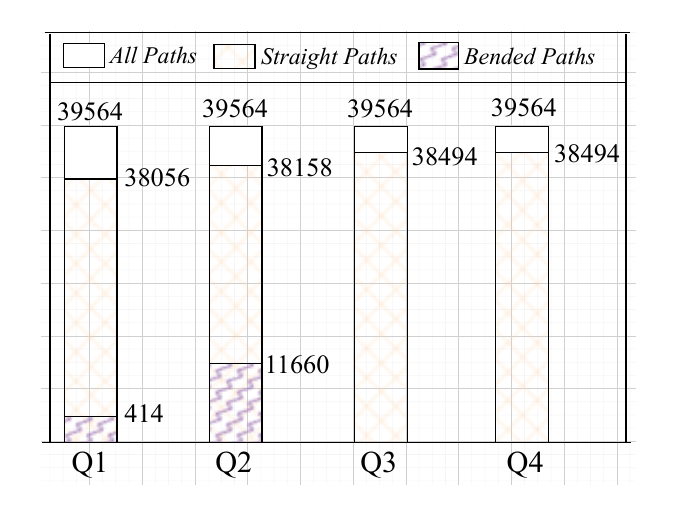}
		\caption{ Inference Performance on Dallas Q2}
	\end{subfigure}
	\begin{subfigure}{0.40\linewidth}
        \centering
        \includegraphics[width=\linewidth]{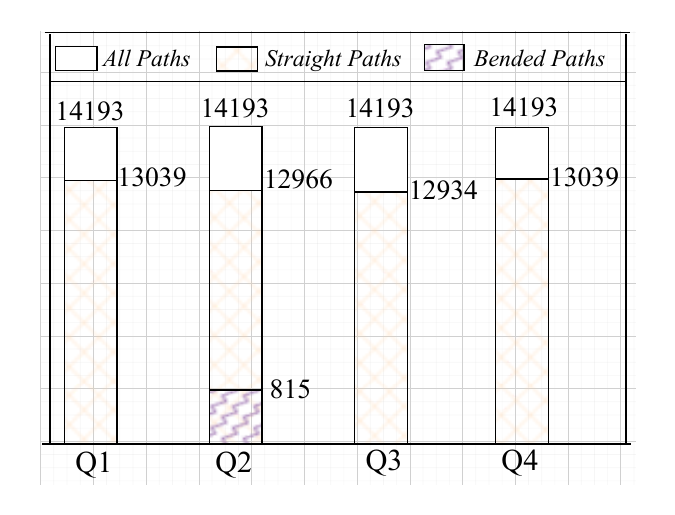}
        		\caption{Inference Performance on Houston Q3}
	\end{subfigure}\vspace{-0.05in}
  \caption{Performance of quadrant-wise inference for Dallas and Houston when the model is trained on one part of the city (i.e., one quadrant) and analyzed on another part. The performance is intuitive to the phenomenon that the best performance is received when the model is trained and inferences on the twin from same quadrant. Each bar represents when the model is trained on that quadrant.}
    \vspace{-0.2in}
	\label{fig:quadrant-wise-inference}
\end{figure}

\noindent $\bullet$ {\bf Downtown Dallas (Inference).} For the inference, we perform various quadrant-wise experiments: (i) inference on same quadrant, the model is trained on, and (ii) inference on different quadrant than the model is trained on. For the case (i), we observe that the trained DQN agent was able to reach to target for most of the time, the detailed plots are in Fig.~\ref{fig:Outdoors-Inference}. For the case (ii), we conduct experiments using the DQN agent trained on all quadrants (Q1, Q2, Q3, and Q4) individually, and infer on the Q2 digital twin, the results are is shown in Fig.~\ref{fig:quadrant-wise-inference} (a). From the plot, it is evident that, during inference, the trained agent on Q2 performed best on Q2, while competitive performance was given by the agent trained on Q1. Intuitively, this is due to the fact of similar landscape of the areas. We observe similar characteristics for other quadrants of the Dallas digital twin. 

\begin{figure}
     \centering    
     \includegraphics[width=1\linewidth]{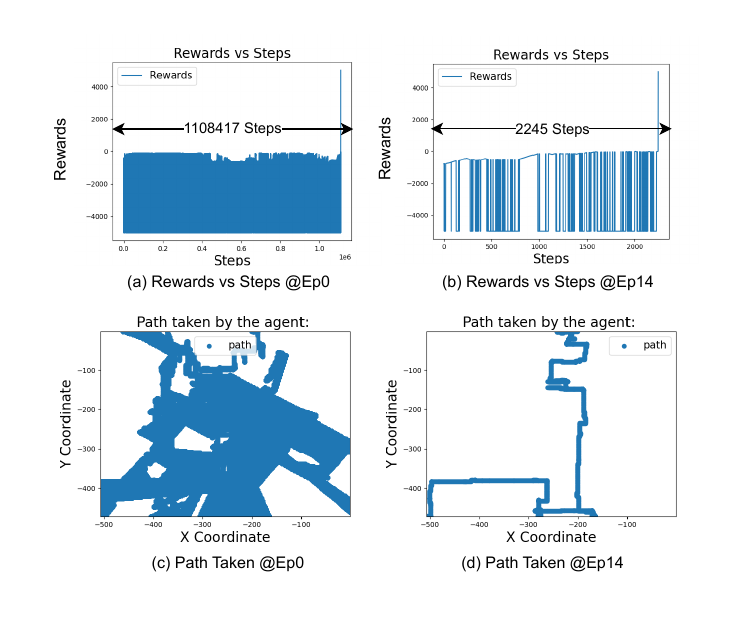}\vspace{-0.25in}
     \caption{The training performance of the $\name$ DQN agent on the blueprint twin of Houston Q1. We see 99.7\% of reduction is required number of steps to 14th episode.}
     \vspace{-0.2in}
     \label{fig:houston-q1-training-plot}
 \end{figure}

\begin{figure}
     \centering    
     \includegraphics[width=1\linewidth]{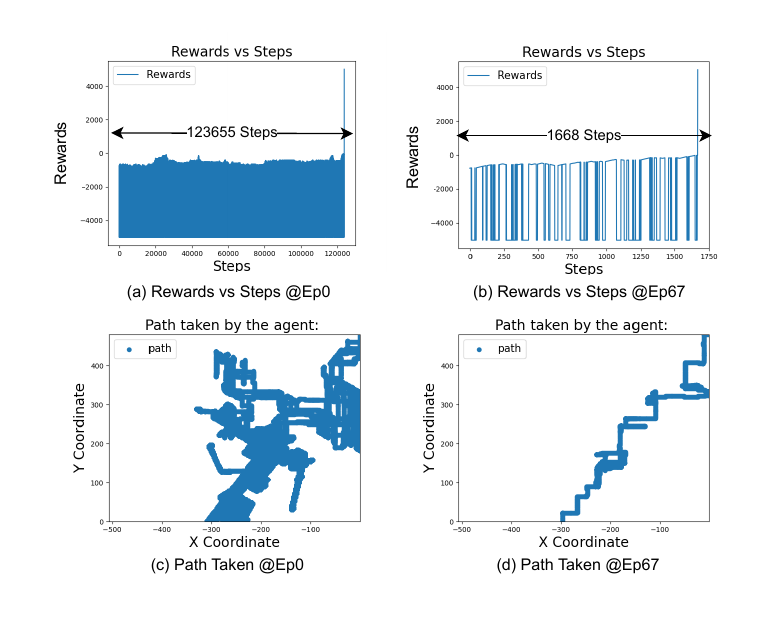}\vspace{-0.25in}
     \caption{The training performance of the $\name$ DQN agent on the blueprint twin of Houston Q2. We see 98.6\% of reduction is required number of steps to 67th episode.}
     \vspace{-0.2in}
     \label{fig:houston-q2-training-plot}
 \end{figure}

\begin{figure}
     \centering    
     \includegraphics[width=1\linewidth]{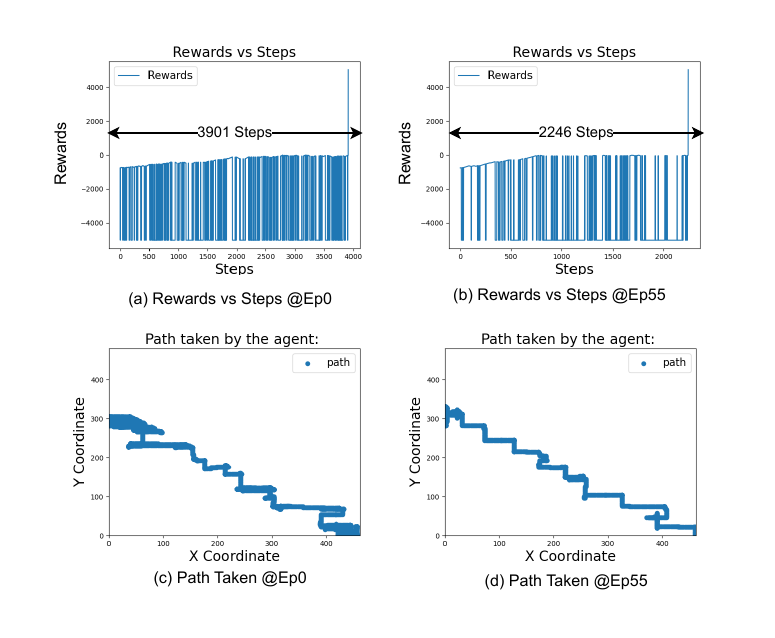}\vspace{-0.25in}
     \caption{The training performance of the $\name$ DQN agent on the blueprint twin of Houston Q3. We see 42.4\% of reduction is required number of steps to 55th episode.}
     \vspace{-0.2in}
     \label{fig:houston-q3-training-plot}
 \end{figure}

\begin{figure}
     \centering   
     \includegraphics[width=1\linewidth]{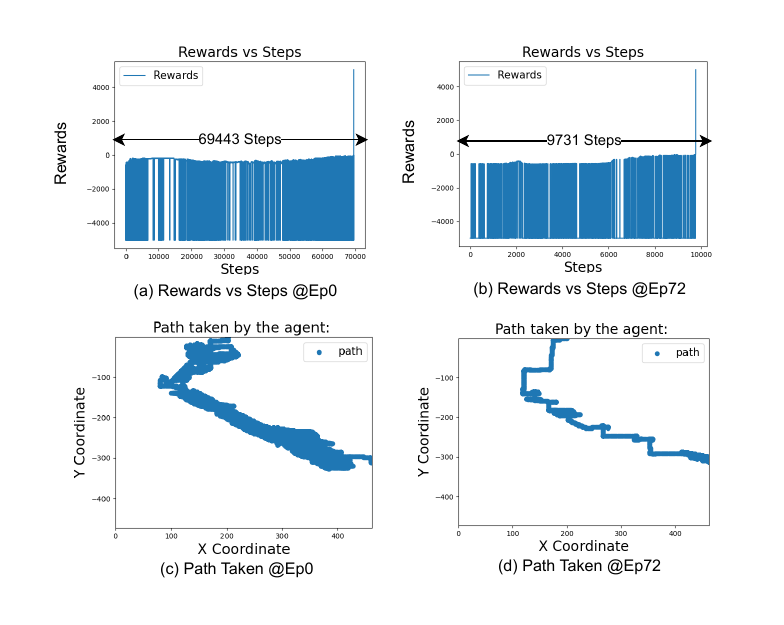}
     \vspace{-0.4in}
     \caption{The training performance of the $\name$ DQN agent on the blueprint twin of Houston Q4. We see 85.98\% of reduction is required number of steps to 72nd episode.}
     \vspace{-0.2in}
     \label{fig:houston-q4-training-plot}
 \end{figure}

\noindent $\bullet$ {\bf Downtown Houston (Training).}
For Houston scenario, we observe, the minimum required episodes for the agent to learn is 14th, 67th, 55th, and 72nd for Houston Q1, Q2, Q3, and Q4 scenarios, respectively. The detailed plots showing the training performance for these experiments are in Figs.~\ref{fig:houston-q1-training-plot}, \ref{fig:houston-q2-training-plot}, \ref{fig:houston-q3-training-plot}, \ref{fig:houston-q4-training-plot} for Houston Q1, Q2, Q3, and Q4 scenarios, respectively. 

\begin{figure}
	\centering
	\begin{subfigure}{0.4\linewidth}
        \centering
        \includegraphics[width=\linewidth]{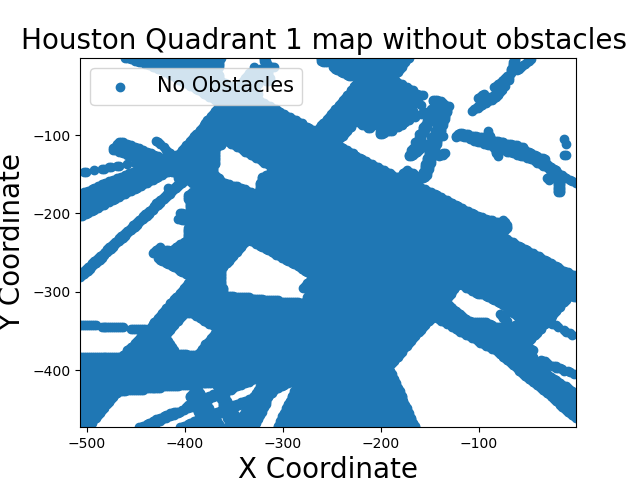}
		\caption{Houston-Q1 Scene}
	\end{subfigure}
	\begin{subfigure}{0.4\linewidth}
        \centering
        \includegraphics[width=\linewidth]{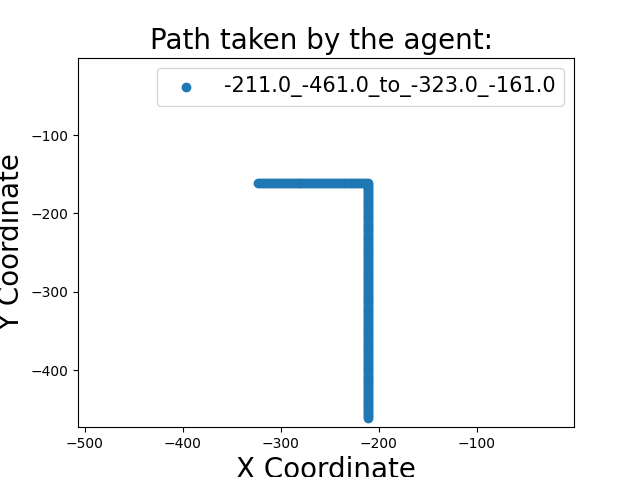}
        		\caption{Houston-Q1 NavPath@Infer}
	\end{subfigure}
        \begin{subfigure}{0.4\linewidth}
        \centering
        \includegraphics[width=\linewidth]{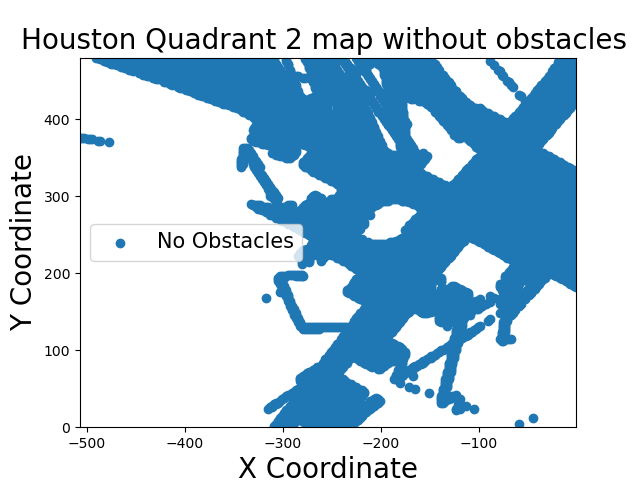}
        		\caption{Houston-Q2 Scene}
	\end{subfigure} 
        \begin{subfigure}{0.4\linewidth}
        \centering
        \includegraphics[width=\linewidth]{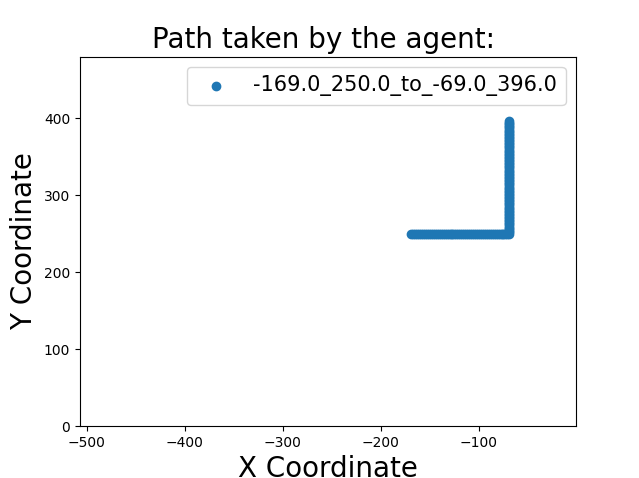}
        		\caption{Houston-Q2 Nav-Path@Infer}
	\end{subfigure} 
 	\begin{subfigure}{0.4\linewidth}
        \centering
        \includegraphics[width=\linewidth]{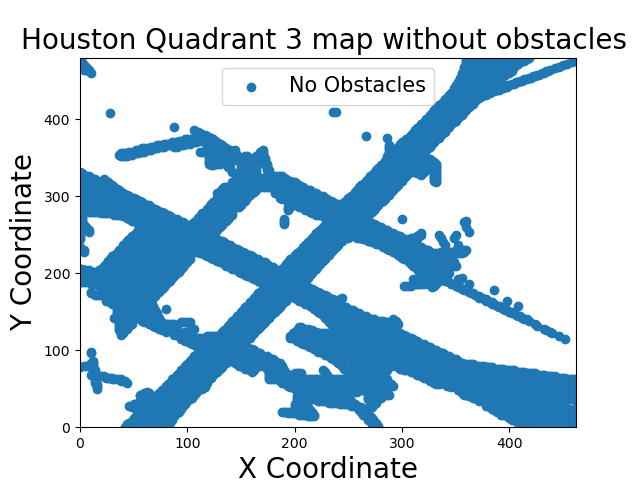}
		\caption{Houston-Q3@Scene}
	\end{subfigure}
	\begin{subfigure}{0.4\linewidth}
        \centering
        \includegraphics[width=\linewidth]{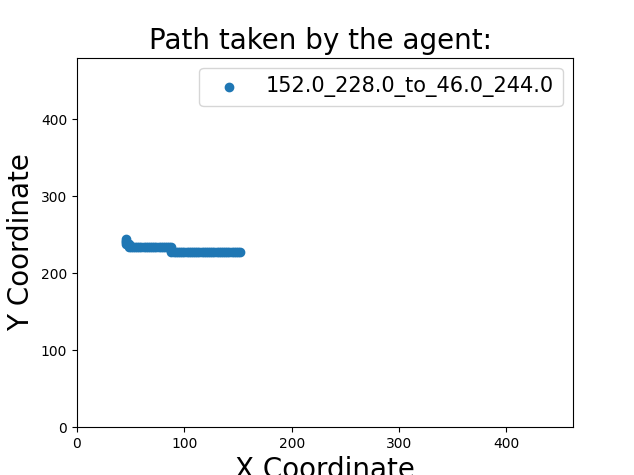}
        		\caption{Houston-Q3 Nav-Path@Infer}
	\end{subfigure} 
 	\begin{subfigure}{0.4\linewidth}
        \centering
        \includegraphics[width=\linewidth]{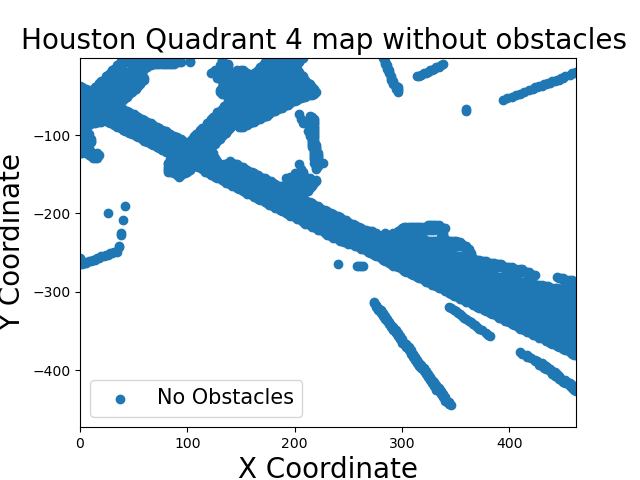}
		\caption{Houston-Q4 Scene}
	\end{subfigure}
	\begin{subfigure}{0.4\linewidth}
        \centering
        \includegraphics[width=\linewidth]{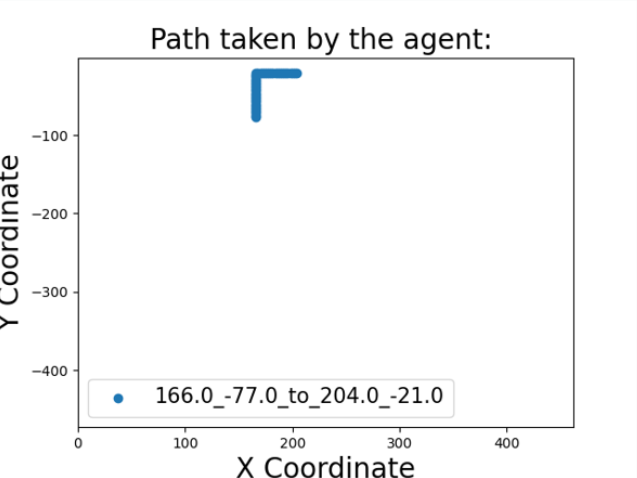}
        		\caption{Houston-Q4 Nav-Path@Infer}
	\end{subfigure}\vspace{-0.05in}
     
  \caption{Inference on blueprint digital twins (downtown Houston). Plots (a), (c), (e), and (g) illustrate the non-obstacle regions within each quadrant. Plots (b), (d), (f), and (h) highlight the robot's successful traversal between points, adjusting its X and Y coordinates and avoiding obstacles in quadrants Q1, Q2, Q3, and Q4.}
    \vspace{-0.5in}
	\label{fig:Outdoors-Inference2}
\end{figure}

\noindent $\bullet$ {\bf Downtown Houston (Inference).}
Similar to the experiments conducted for Dallas, we also conduct quadrant-based testing for Houston. This involves: (i) running inference on the same quadrant the model was trained on, and (ii) running inference on a different quadrant from the one the model was trained on. In case (i), we observe that the trained DQN agent reaches the target almost every time successfully, with detailed plots presented in Fig.~\ref{fig:Outdoors-Inference2}. In case (ii), we use the DQN agent trained in Q2 and evaluate its performance across all quadrants, with the results illustrated in Fig.~\ref{fig:quadrant-wise-inference} (b). The plot shows interesting behavior that the trained agent on Q2 performs best in Q3 digital twin of Houston. \vspace{-0.05in}


\begin{observation}
Training: We observe that the trained $\name$ DQN agent learns the optimal path by avoiding obstacles and minimizes the collisions for all four blueprint digital twins (see Figs.~\ref{fig:lab-cubicle-training-plot}, \ref{fig:lab-meeting-training-plot}, \ref{fig:dallas-q1-training-plot}, \ref{fig:dallas-q2-training-plot}, \ref{fig:dallas-q3-training-plot}, \ref{fig:dallas-q4-training-plot}, \ref{fig:houston-q1-training-plot}, \ref{fig:houston-q2-training-plot}, \ref{fig:houston-q3-training-plot}, \ref{fig:houston-q4-training-plot}).\vspace{-0.08in} 
\end{observation}

\begin{observation}
Inference: We observe that the generated navigation paths by the $\name$ DQN agent reaches the target location by avoiding the  obstacle regions of various indoor and outdoor scenarios (see Figs.~\ref{fig:Cubicle-Meeting Inference},~\ref{fig:Outdoors-Inference},~\ref{fig:Outdoors-Inference2}). \vspace{-0.08in}
\end{observation}

\subsubsection{Minimizing the Collisions in $\name$}
\label{subsec:collisions}
During training, the agent explores the environment and initially experiences a large number of negative rewards due to frequent collisions. However, over time, it learns to minimize these collisions by avoiding obstacles and improving its overall training performance for all blueprint twins (Figs.~\ref{fig:lab-cubicle-training-plot}, \ref{fig:lab-meeting-training-plot}, \ref{fig:dallas-q1-training-plot}, \ref{fig:dallas-q2-training-plot}, \ref{fig:dallas-q3-training-plot}, \ref{fig:dallas-q4-training-plot}, \ref{fig:houston-q1-training-plot}, \ref{fig:houston-q2-training-plot}, \ref{fig:houston-q3-training-plot}, and \ref{fig:houston-q4-training-plot})  illustrates the agent's collision frequency during training and how it gradually improved, as reflected in the increase in rewards and the reduction in steps taken. After training, the agent successfully learned the optimal path to reach its target from the starting point while avoiding obstacles as much as possible, as demonstrated in Figs.~\ref{fig:lab-cubicle-training-plot}, \ref{fig:lab-meeting-training-plot}, \ref{fig:dallas-q1-training-plot}, \ref{fig:dallas-q2-training-plot}, \ref{fig:dallas-q3-training-plot}, \ref{fig:dallas-q4-training-plot}, \ref{fig:houston-q1-training-plot}, \ref{fig:houston-q2-training-plot}, \ref{fig:houston-q3-training-plot}, and \ref{fig:houston-q4-training-plot}. Table~\ref{tab:Min-Steps} presents the scene description, the number of steps taken in the episode, and the X and Y coordinates from the source to the destination for training in all the blueprint twins.

\begin{table}
\centering
 \caption{Glimpse of the training performance for all the blueprint twins w.r.t. minimum steps taken. The X and Y coordinates are relative locations received from the robot odometer sensor.}
\resizebox{0.9\linewidth}{!}{
\begin{tabular}{|c|c|c|}
\hline
\textbf{Scene Description}  & \textbf{ X and Y Coordinates} &   \textbf{Minimum Steps Taken}\\
\hline
\hline
Lab Cubicle & [-8, -9] [17, 8] & 45  in 13 Episode\\
\hline
Lab Meeting & [-20, -22] [17, 17] & 78  in 389 Episode\\
\hline
Dallas Q1 & [-275, -328] [-175, -1] & 808  in 19 Episode\\
\hline
Dallas Q2 & [-160, 0] [-200, 360] & 1475 in 32 Episode\\
\hline
Dallas Q3 & [200, 0] [0, 160] & 416 in 37 Episode\\
\hline
Dallas Q4 & [75, -328] [0, -160] & 307 in 25 Episode\\
\hline
Houston Q1 & [-570, -473] [-1, -161] & 2245 in 14 Episode\\
\hline
Houston Q2 & [-297, 0] [-1, 140] & 1668 in 67 Episode\\
\hline
Houston Q3 & [460, 0] [0, 300] & 2246 in 55 Episode\\
\hline
Houston Q4 & [-462, -301] [200, -1] & 6431 in 72 Episode\\
\hline
\end{tabular}}
\vspace{-0.1in}
 \label{tab:Min-Steps}
\end{table}

\subsection{Performance of $\name$ DQN Agent (Trained on Blueprint Twin and Inference on Dynamic Twin)}\vspace{-0.05in} In the second set of experiments, we explore the performance of  $\name$ DQN agent when it is trained on a specific digital twin and later the digital twin is updated with more (or less) objects within the scene. The details about the dynamic digital twin generation is discussed in Sec.~\ref{subsubsec:testbed_dynamic_twin}.

\begin{figure}
	\centering
	\begin{subfigure}{0.4\linewidth}
        \centering
        \includegraphics[width=\linewidth]{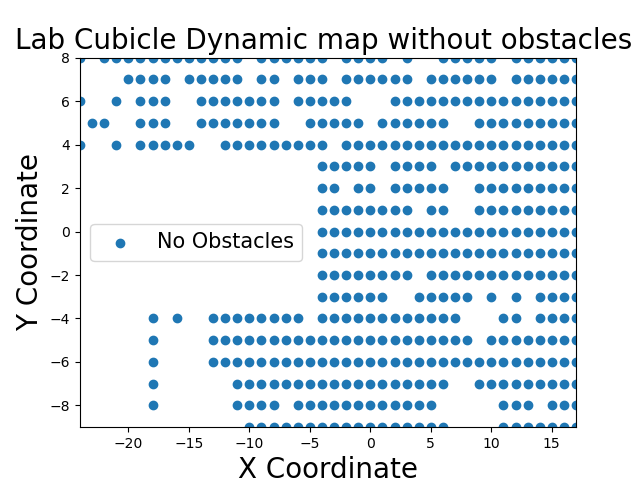}
		\caption{Cubicle Scene}
	\end{subfigure}
	\begin{subfigure}{0.4\linewidth}
        \centering
        \includegraphics[width=\linewidth]{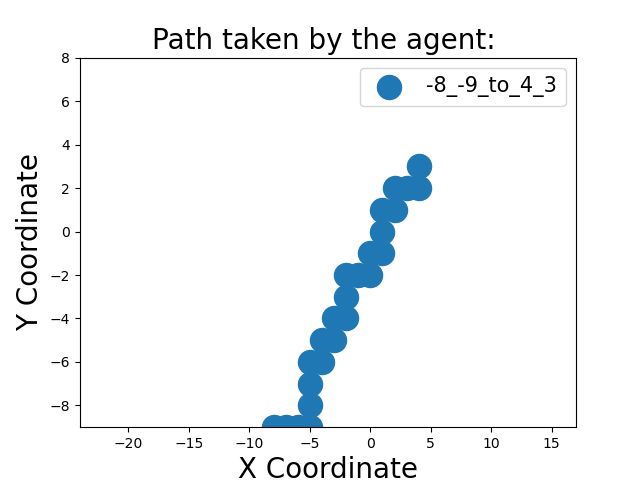}
        		\caption{Nav Path@Cubicle}
	\end{subfigure} 
 	\begin{subfigure}{0.4\linewidth}
        \centering
        \includegraphics[width=\linewidth]{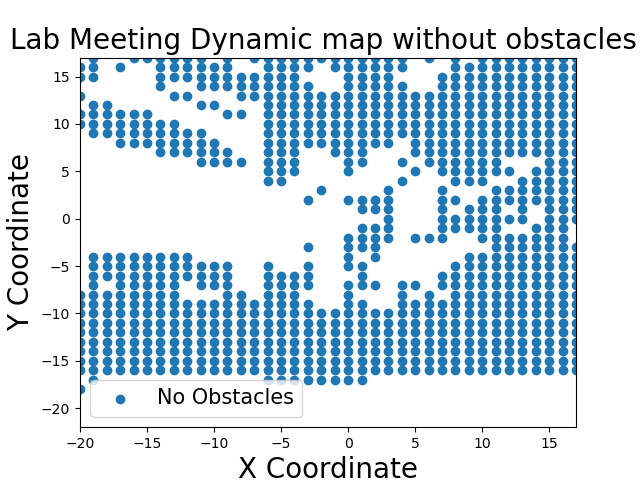}
		\caption{Meeting Scene}
	\end{subfigure}
        \begin{subfigure}{0.4\linewidth}
        \centering
        \includegraphics[width=\linewidth]{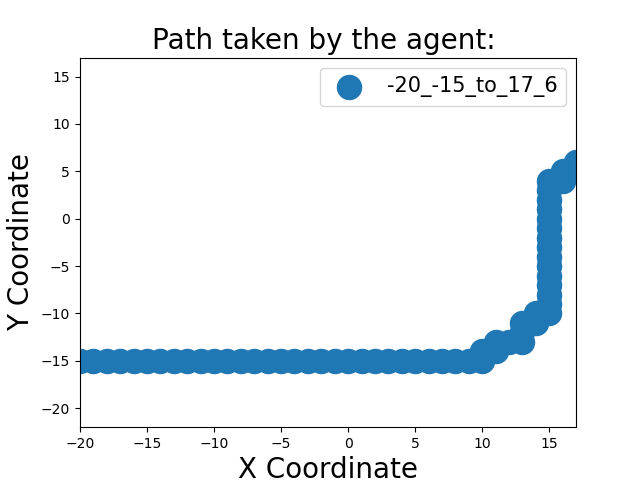}
        
		\caption{Nav Path@Meeting}
	\end{subfigure}\vspace{-0.05in}
  \caption{Inference on dynamic digital twins (lab cubicle and meeting space). In (a) and (c), the areas with and without obstacles are shown for the cubicle and meeting space scenes, respectively. While (b) and (d) depict the robot's successful navigation, avoiding obstacles, for cubicle and meeting spaces' dynamic digital twins, respectively.}
    \vspace{-0.2in}
	\label{fig:Indoors-Inference-dynamic}
\end{figure}

\noindent $\bullet$ {\bf Dynamic Lab Cubicle (Inference).} The DQN agent which is trained on the blueprint lab cubicle digital twin is analyzed on the dynamic lab cubicle digital twin, the plots are shown in Fig.~\ref{fig:Indoors-Inference-dynamic} (a) and (b). The blue dots in Fig.~\ref{fig:Indoors-Inference-dynamic} (a) shows no obstacle regions or open areas within the dynamic digital twin of cubicle, which is different from the map shown in blueprint twin of the same in Fig.~\ref{fig:Cubicle-Meeting Inference} (a). The navigation path shown in Fig.~\ref{fig:Indoors-Inference-dynamic} (b) shows that the trained DQN agent is able to successfully navigate even with the dynamic changes. 

\noindent $\bullet$ {\bf Dynamic Meeting Space (Inference).} Similarly, the trained DQN agent on the blueprint lab meeting digital twin is analyzed on the dynamic lab meeting digital twin, the plots are in Fig.~\ref{fig:Indoors-Inference-dynamic} (b) and (d). Fig.~\ref{fig:Indoors-Inference-dynamic} (a) shows no obstacle regions or open areas within the dynamic digital twin of meeting space (the blueprint twin of the same is in Fig.~\ref{fig:Cubicle-Meeting Inference} (c)). The navigation path is shown in Fig.~\ref{fig:Indoors-Inference-dynamic} (d).

\begin{figure}
	\centering
	\begin{subfigure}{0.40\linewidth}
        \centering
        \includegraphics[width=\linewidth]{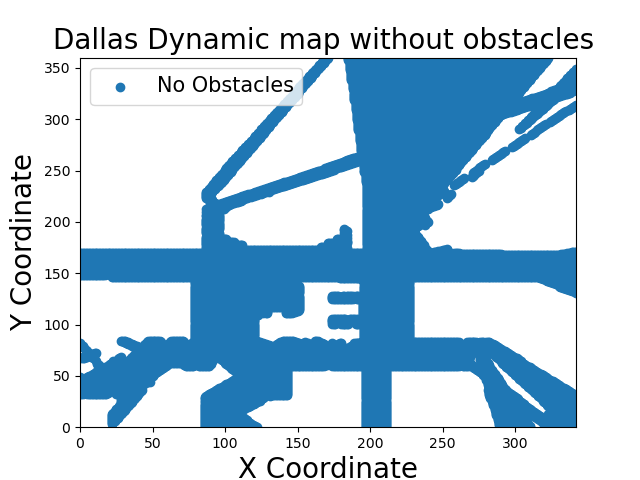}
		\caption{Dallas Q2 Scene}
	\end{subfigure}
	\begin{subfigure}{0.4\linewidth}
        \centering
        \includegraphics[width=\linewidth]{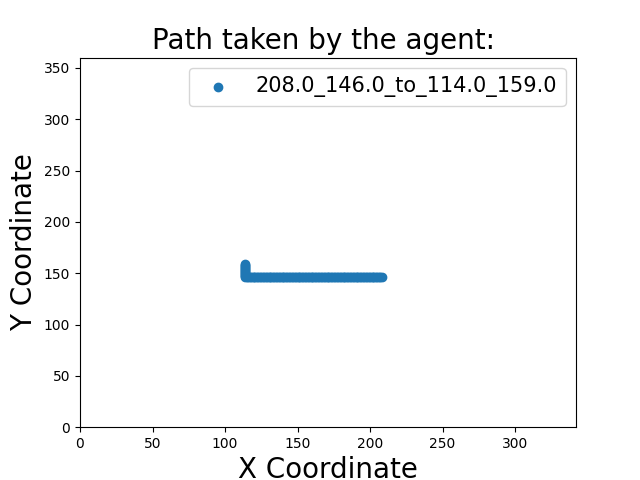}
        		\caption{Dallas Q2 Inference}
	\end{subfigure}
        \centering
	\begin{subfigure}{0.4\linewidth}
        \centering
        \includegraphics[width=\linewidth]{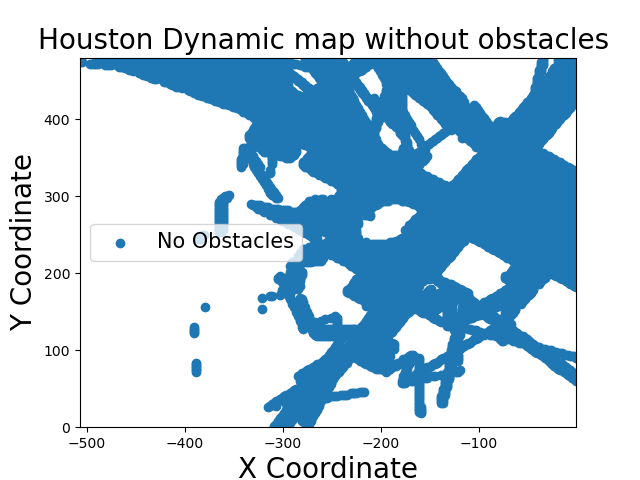}
		\caption{Houston Q3 Scene}
	\end{subfigure}
	\begin{subfigure}{0.4\linewidth}
        \centering
        \includegraphics[width=\linewidth]{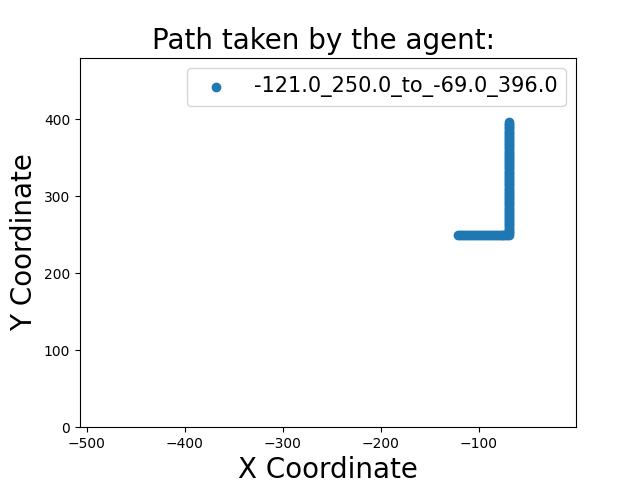}
        \caption{Houston Q3 Inference}
	\end{subfigure} 
    \vspace{-0.05in}
  \caption{Inference on dynamic digital twins (Dallas and Houston). In (a) and (c), the areas with and without blockages and obstacles are shown for the Dallas and Houston downtown, respectively. The paths in (b) and (d) depict the robot's successful navigation, avoiding blockages and obstacles.}
    \vspace{-0.2in}
	\label{fig:outdoor-dynamic}
\end{figure}

\noindent $\bullet$ {\bf Dynamic Downtown Dallas (Inference).} For the outdoor scenarios, the trained model on the blueprint twins are analyzed over the dynamic twin for Dallas digital twin. The plots to signify the non-blocked areas and a sample navigation path are given in Fig.~\ref{fig:outdoor-dynamic} (a) and (b), respectively. Note that generated map signifying the non-blocked areas (i.e., where there are no buildings) in Fig.~\ref{fig:outdoor-dynamic} (a) is different than the blueprint twin of the same shown in Fig.~\ref{fig:Outdoors-Inference} (e). 

\noindent $\bullet$ {\bf Dynamic Downtown Houston (Inference).}  The plots for non-blocked areas and a sample navigation path for the Houston are given in Fig.~\ref{fig:outdoor-dynamic} (c) and (d), respectively. Similar to the Dallas case, for Houston also the generated maps signifying the non-blocked areas in Fig.~\ref{fig:outdoor-dynamic} (c) is different than the blueprint twin of the same shown in Fig.~\ref{fig:Outdoors-Inference2} (c).

\begin{observation}
    The trained DQN agent is able to successfully navigate within the dynamic changes (see Figs.~\ref{fig:Indoors-Inference-dynamic} and ~\ref{fig:outdoor-dynamic}).\vspace{-0.05in}
\end{observation}

\begin{remark}
    The in-depth analysis of the percentage of changes a trained model can adapt, is open for further investigation.
    \vspace{-0.1in}
\end{remark}

\subsection{Comparison with State-of-the-art}\vspace{-0.05in}
\label{subSec:sota_compare}
We compare the overall training time with the state-of-the-art~\cite{niu2021accelerated, beomsoo2021mobile, hu2020voronoi}, it is evident in 
Table.~\ref{tab:sota_comparision}, that $\name$ outperforms the other methods in terms of training time even when trained on a computing platform with only CPU.\vspace{-0.05in}

\begin{table}[!t]
\centering
 \caption{Comparison with state-of-the-art w.r.t. training time.}
\resizebox{1.0\linewidth}{!}{
\begin{tabular}{|c|c|c|p{1.2in}|c|}
\hline
\textbf{Paper}  & \textbf{Steps} &  \textbf{Sensors} & \textbf{Computing} & \textbf{Training}\\
 & &  & \textbf{Platform}& \textbf{Time}\\
\hline
\hline
Niu {\em et al.}~\cite{niu2021accelerated} & 25K& LiDAR & NVIDIA TITAN RTX GPU& 133 minutes\\
\hline
 Beomsoo {\em et al.}~\cite{beomsoo2021mobile} & 600K&  LiDAR &  Not mentioned & 180 minutes\\
\hline
 Hu {\em et al.}~\cite{hu2020voronoi}& 10K & LiDAR & NVIDIA GTX 1080 GPU and Intel Core i9 CPU& 40 minutes\\
 \hline
\hline
$\name$& 600K &  Ray-tracing Data & 32 GB Ryzen CPU& 46 minutes\\
\hline
\end{tabular}}
\vspace{-0.2in}
 \label{tab:sota_comparision}
\end{table}

\begin{observation}
    We observe that the proposed $\name$ framework is able to yield competitive training time with the state-of-the-art approaches. $\name$ requires lesser training time ($\sim < 74.4\%$) even when run on a CPU computing platform (see Table~\ref{tab:sota_comparision}, validates Contribution 5). 
    \vspace{-0.05in}
\end{observation}

\subsection{End-to-end Decision Delay}
\label{subSec:delay}\vspace{-0.03in}
The end-to-end delay is calculated based on the two measurements: (a) the ray-tracing time per step, which is $16ms$ and (b) the inference time to generate the next action for the robot, which is $17ms$. Overall, the end-to-end decision delay is summed up to be: $33ms$ when ran on an edge device.
\vspace{-0.05in} 


\subsection{Applicability of $\name$}
\vspace{-0.03in}
The $\name$ framework is particularly applicable in privacy-sensitive environments, such as hospitals. In these dynamic settings, where doctors, patients, and visitors are constantly moving, traditional SLAM algorithms struggle with navigation challenges~\cite{slam_dynamic}. $\name$ enables robots to navigate effectively in crowded spaces while protecting patient privacy. As shown in Fig.~\ref{fig:applicability}, a robot can autonomously reach a patient's room to initiate a teleconference for doctor-patient communication, utilizing ray-tracing data to avoid potential privacy risks associated with conventional sensors. This approach ensures efficient navigation while addressing the privacy concerns of all individuals present. Next we discuss, how the $\name$ can be used in such scenario:

 \begin{figure}[t!]
     \centering    
     \includegraphics[width=0.8\linewidth]{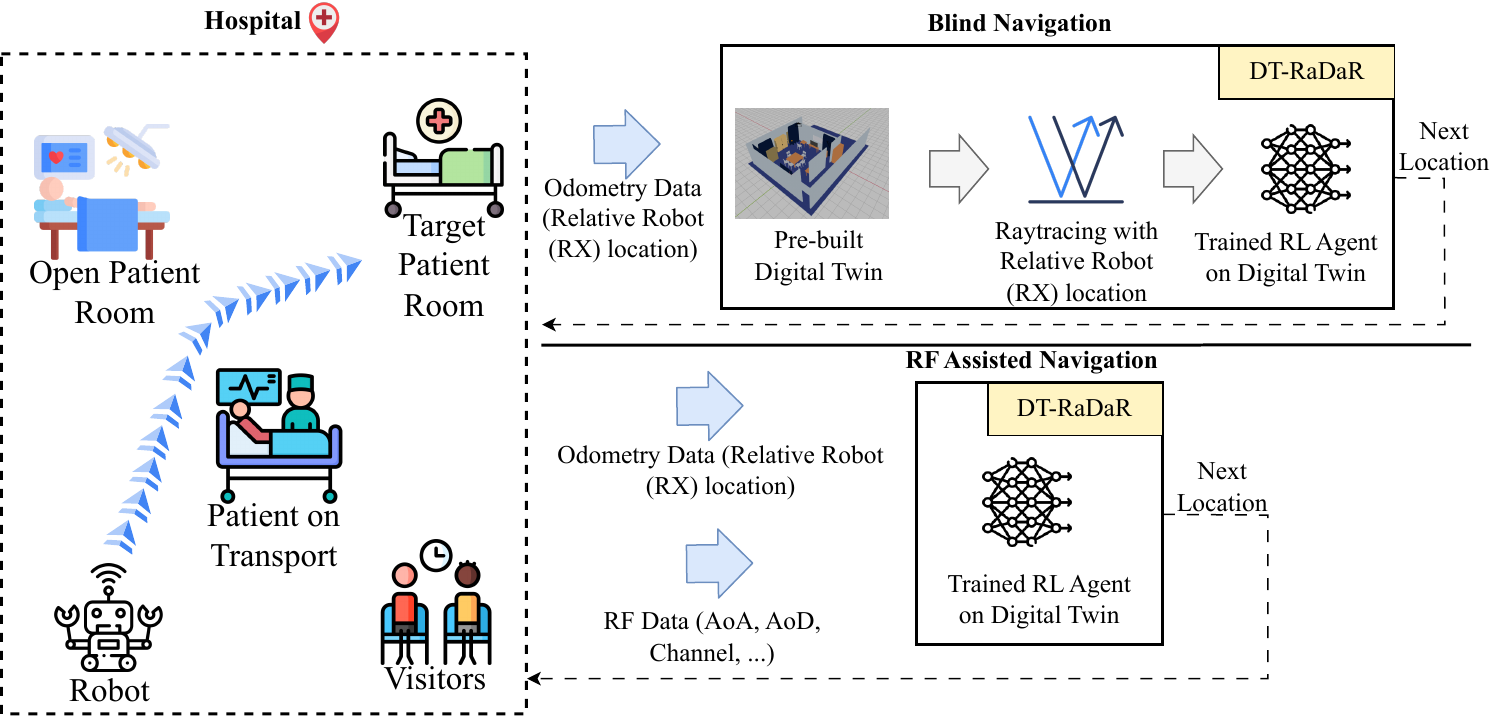}
     \caption{Applicability of $\name$ in a hospital scenario. We show two approaches on how $\name$ can be used for robot navigation in the hospital scenario: (a) blind and (b) RF-assisted navigation.}
     \vspace{-0.2in}
     \label{fig:applicability}
 \end{figure}

\subsubsection{$\name$ in Blind Navigation}
One way of using $\name$ is blind navigation strategy where only the robot's odometery data is fed to our framework as shown in Fig.~\ref{fig:applicability}. The robot will be using off-the-shelf $\name$ framework which includes a pre-build digital twin, ray-tracing suite, and trained RL model. However, this approach may result in suboptimal performance if the real-world environment undergoes significant changes that differ from the digital twin model.

\subsubsection{$\name$ in RF-assisted Navigation}
Various RF data such as angle-of-arrival, angle-of-departure, and channel coefficients~\cite{Yang_2021, Shih_2023} are fed to the trained RL agent of $\name$, as shown in Fig.~\ref{fig:applicability}. The RL agent is trained on a digital twin of the hospital scenario, enabling real-time navigation with RF data for accuracy and privacy preservation.

\subsection{Limitations and Future Possibilities of $\name$}
\vspace{-0.03in}
\label{subsnippiHosec:limitation}
The $\name$ framework establishes a foundational approach for privacy-preserving robot navigation using ray-tracing data, but it has limitations that highlight future research opportunities. These include potential discrepancies between real-world environments and digital twin representations, the impact of external factors like weather on signal quality, and the current inability of digital twins to capture fine-grained, rapid changes in the physical world. Additionally, while the framework's privacy benefits are a key contribution, further emphasis on existing work addressing privacy concerns in sensor-based navigation is needed to clearly demonstrate $\name$'s superior protection. Addressing these challenges could involve implementing dynamic update loops between digital twins and real environments, developing robust interference mitigation strategies, identifying optimal application scenarios, and conducting comparative privacy analyses against traditional sensor-based methods.
\vspace*{-3pt}

\section{Conclusion}\vspace{-0.05in}
\label{sec:conclusions}
We propose a DQN based robot navigation approach which is applied on the digital twins of a real-world scenario for the navigation of delivery or service robots. The digital twins are designed using open source Blender software and the ray-tracing data is generated through the Sionna RT software. The thorough experimentation shows that the idea of using ray tracing data for robot navigation is feasible. It also provides some resiliency when trained over the digital twins of one scenario and analyzed over the digital twins of another type of scenario. The experimentation on quantifying such resiliency and extending to more generalized twins would be our next step for future work.

\vspace*{-3pt}

\vspace*{-3pt}
\section*{Acknowledgment}\vspace{-0.05in}
\vspace*{-3pt}
We acknowledge usage of AI tools for solely language polishing purpose only. 

\bibliographystyle{IEEEtran} 
\bibliography{reference}

\end{document}